\documentclass{aa}  
\usepackage{graphicx}
\usepackage{epsfig}
\usepackage{txfonts}
\begin{document}
   \title{The evolution of the mass-metallicity relation 
in galaxies of different morphological types}

   \author{F. Calura$^1$\thanks{fcalura@oats.inaf.it}, A. Pipino$^2$, C. Chiappini$^{3,4}$, F. Matteucci$^{1,4}$, R. Maiolino$^5$}
 
   \offprints{F. Calura}

   \institute{
	      $1$ Dipartimento di Astronomia, Universit\'a di Trieste, 
		Via G.B. Tiepolo 11, 34143 Trieste, Italy\\
              $2$ Department of Physics \& Astronomy , University of Southern California, Los Angeles 90089-0484, USA\\
              $3$ Observatoire de Gen\`eve, Universit\`e de Gen\`eve, 51 Chemin des Maillettes, CH-1290 Sauverny, Switzerland\\
              $4$ INAF- Osservatorio Astronomico di Trieste, Via G. B. Tiepolo 11, 34131 Trieste, Italy\\
              $5$ INAF - Osservatorio Astronomico di Roma, via di Frascati 33, 00040 Monte Porzio Catone, Italy 
        }

   \date{Received; accepted}

 
  \abstract
   {}
{By means of chemical evolution models for ellipticals, spirals and
irregular galaxies, we aim at investigating the physical meaning
and the redshift evolution of the mass-metallicity relation as well
as how this relation is connected with galaxy morphology.}  
{Our models distinguish among different morphological
types through the use of different infall, outflow, and star
formation prescriptions.  We assume that galaxy morphologies do not
change with cosmic time.  We present a method to account for a
spread in the epochs of galaxy formation and to refine the galactic
mass grid.  To do that, we extract the formation times randomly and
assume an age dispersion $\Delta_t$.  We compare our predictions to
observational results obtained for galaxies between redshifts 0.07
and 3.5.}  {We reproduce the mass-metallicity (MZ) relation mainly
by means of an increasing efficiency of star formation with mass in
galaxies of all morphological types, without any need to 
invokegalactic outflows favoring the loss of metals in the less massive
galaxies. Our predictions 
can help constraining the slope and the zero point of the
observed local MZ relation, both affected by uncertainties
related to the use of different metallicity calibrations.  We show
how, by considering the MZ, the O/H vs  star formation rate (SFR),
and the SFR vs galactic mass diagrams at various redshifts, it is
possible to constrain the morphology of the galaxies producing
these relations.  
Our results indicate that the galaxies observed
at $z=3.5$ should be mainly proto-ellipticals, whereas 
at $z=2.2$ the observed galaxies consist
of a morphological mix of proto-spirals and proto-ellipticals. 
At lower redshifts, the observed MZ relation is
well reproduced by considering both spirals and irregulars. 
Galaxies with different star formation histories
may  overlap in the MZ diagram, but measures of abundance ratios such
as [O/Fe] can help to break this degeneracy.  Predictions for the
MZ relations for other elements (C, N, Mg, Si, Fe) are also
presented, with  largest dispersions predicted for elements
produced in considerable amounts by Type Ia SNe, owing to the long lifetimes of their progenitors.} 
   {}
\keywords{galaxies: abundances; ISM: abundances; galaxies:high-redshift; galaxies: evolution. }
  \authorrunning{Calura et al.}
  \titlerunning{The evolution of the MZ relation in galaxies of different morphological types}
 
\maketitle
%
\section{Introduction}
In the last few years, several observational projects have been devoted 
to the study of the relation between galactic stellar mass and metallicity 
in star forming galaxies. 
In particular, the study of the redshift evolution of the mass-metallicity (hereinafter MZ) relation 
has provided us with crucial information 
on the cosmic evolution of star formation, as well as on 
the temporal evolution of the chemical properties 
of the stellar populations. 
Various observational  studies of the MZ
relation have outlined that across a large redshift 
range, i.e. between z=3.5 and z=0, 
in the high-mass, high-metallicity part of the MZ plot,
galaxies tend to form a plateau, whereas at smaller masses  
the MZ plot shows an increase of metallicity with mass. 
Moreover, both the zero point and 
the slope of the MZ relation 
are a function of redshift, as shown in a recent paper by Maiolino et al. (2008).
Therefore, the metallicity evolution is stronger for low mass galaxies 
than for high mass galaxies. \\
Various theoretical explanations of the MZ relation have been proposed so far. 
One interpretation is linked to starburst-induced galactic outflows, 
more efficient in expelling metal-enriched gas in dwarf galaxies than in giant galaxies, 
owing to the shallow gravitational potential wells of the former (Larson 1974; Dekel \& Silk 1986; Tremonti et al. 2004, De Lucia et al. 2004; 
Kobayashi, Springel \& White  2007,  
Finlator \& Dav\'e 2008). 
Alternatively, the dilution caused by infall can act in a way similar to the presence of outflows, 
once one assumes longer infall times in lower mass galaxies (Dalcanton et al. 2004). 
Another possibility is that low mass galaxies are less evolved than 
large galaxies. In this picture, 
while large galaxies have formed the bulk of their stars by means of 
an intense event at high redshift, enriching quickly their ISM to solar or over-solar metallicities, 
dwarf galaxies, characterized by lower star formation efficiencies (i.e. star formation rates per unit mass of gas) have sub-solar interstellar metallicities. 
This interpretation is supported by several chemical evolution models (Lequeux et al. 1979, Matteucci 1994), by cosmological N-body 
simulations (Brooks et al. 2007; Mouhcine et al. 2008, Tassis et al. 2008) 
and by hydrodynamical simulations (Tissera et al. 2005, De Rossi et al. 2007,). 
A third interpretation of the MZ relation is linked to the initial mass function. 
K\"oppen et al. (2007) showed how the MZ relation can be explained by 
a higher upper mass cutoff in the initial mass function (IMF) in more massive galaxies. \\
However, none of these theoretical studies have investigated 
how the MZ relation is connected with galaxy morphology, 
as well as the role played by the star formation histories of galaxies 
of different morphological types.  

In this paper, our aim is to investigate these aspects, as well as to understand the 
physical meaning of the MZ relation for different galaxies. 
To perform this task, we use detailed chemical evolution models 
for elliptical, spiral and irregular galaxies. 
These models have proven to well reproduce the main chemical properties of elliptical (e.g. Pipino \& Matteucci 2004), spirals (Chiappini et al. 2001, 2003; Cescutti et al. 2007) and irregular galaxies (e.g. Lanfranchi \& Matteucci, 2003), as well as other observational constraints, such as the gas-to-light ratios, the supernova rates as a function of morphological type (Calura \& Matteucci 2006a) and the cosmic evolution of the luminosity density (Calura \& Matteucci 2003; Calura, Matteucci \& Menci 2004).

This paper is organized as follows. In Section 2, we present our models and our methods to 
investigate the MZ relation. In section 3, we describe the set of observational data used 
in this work and we present our results. Finally, in Section 4 we draw some conclusions. 
Throughout the paper, we adopt a concordance $\Lambda$-Cold Dark Matter cosmology, characterized by 
$\Omega_m=0.3$, $\Omega_\Lambda=0.7$ and $h=0.7$. 

\section{Chemical Evolution Models} 
\label{models}

In this paper, we use  chemical evolution 
models for elliptical, spiral and irregular galaxies that have already been tested 
against various local observational constraints. In general, in chemical evolution models, 
the time-evolution of the fractional mass of the element $i$ 
in the gas within a galaxy, $G_{i}$, is described by the basic 
equation:

\begin{equation}
\dot{G_{i}}=-\psi(t)X_{i}(t) + R_{i}(t) + (\dot{G_{i}})_{inf} -
(\dot{G_{i}})_{out}
\end{equation}

\noindent
where $G_{i}(t)=M_{g}(t)X_{i}(t)/M_{tot}$ is the gas mass in 
the form of an element $i$ normalized to the total baryonic mass 
$M_{tot}$ and $G(t)= M_{g}(t)/M_{tot}$ is the total fractional 
mass of gas present in the galaxy at the time $t$. 
The quantity $X_{i}(t)=G_{i}(t)/G(t)$ is the 
abundance by mass of an element $i$, with
the summation over all elements in the gas mixture being equal 
to unity. $\psi(t)$ is the star formation rate (SFR), namely the 
fractional amount of gas turning into 
stars per unit time.
$R_{i}(t)$ is the returned fraction of matter in the 
form of an element $i$ (both newly produced and already present in the stars) that the stars restore into the ISM through 
stellar winds, Type Ia and Type II supernova explosions.
$(\dot{G_{i}})_{inf}$ and  $(\dot{G_{i}})_{out}$ describe   
the possible infall of external gas and the possible presence of outflows, respectively.\\
The nucleosynthesis prescriptions are common to all models. 
For massive stars and Type Ia SNe, we adopt the empirical yields suggested by Fran\c cois et al. (2004),   
which are based on Woosley \& Weaver (1995) for massive stars and on the Type Ia SNe yields of Iwamoto et al. (1999).  
Fow low and intermediate mass stars, the adopted prescriptions are the ones by van den Hoeck $\&$ Groenewegen (1997). The Type Ia SN rate computation is based on the single-degenerate model and follows the Matteucci \& Recchi (2001) prescriptions.  \\
The SFR $\psi(t)$ is a 
Schmidt (1959) law expressed as:

\begin{equation}
\psi(t) = \nu G^{k}(t)
\end{equation}

The quantity $\nu$ is the efficiency of star formation, 
namely the inverse of the typical time-scale for star formation,
and is expressed in $Gyr^{-1}$.\\
Unless otherwise stated, the rate of gas infall, for a given element i, is defined as:
\begin{equation}
(\dot G_{i})_{inf}\,=\, \frac{C_{inf}}{M_{tot}}e^{-t/ \tau}
\end{equation}
\noindent
with $C_{inf}$ being a suitable constant, tuned in order to reproduce the present-day stellar mass,
and $\tau$ the infall timescale.\\
In all our models, the instantaneous recycling approximation (IRA) is relaxed. 
This means that the chemical abundances are computed by taking into account 
the stellar lifetimes. \\
In this work, we assume that interstellar abundances are not affected by dust depletion. 
This assumption is motivated by the fact that the observational abundances considered here are measured 
in $H_{II}$ regions, where dust grains are destroyed by intense the UV radiation fields generated by massive stars 
(Okada et al. 2008). 
Quantitative estimates of the effects of dust on metallicity measures in $H_{II}$ regions indicate that 
the presence of dust grains does not introduce large errors in the global metallicity indicators being 
the uncertainty in the metallicity due to dust effects in any case $\le 0.2$ dex (Shields \& Kennicutt 1996). 
Furthermore, local depletion measurements in different environments indicate that O should not be considered as a strongly   
refractory element (Jensen et al. 2005) . This provides further support to  our assumption of neglecting dust depletion effects in MZ studies. \\
We assume
that galaxy morphology does not change with redshift.  
When discussing different galaxy types
at high redshift, it may seem inappropriate to use the classification
in terms of ellipticals, spirals and irregulars, as we observe
them today.
For this reason, when discussing our results at any redshift, whenever we use to the expression ``ellipticals'', ``spirals'' or ``irregulars'', 
we refer to the high-redshift counterparts of each morphological type, or more appropriately to proto-ellipticals, proto-spirals 
and proto-irregulars.

\subsection{Elliptical galaxies}

For the chemical evolution of elliptical galaxies,  we adopt the one-zone model of Pipino \& 
Matteucci (2004) where we address the reader for further details. 
Here we assume that all spheroids, i.e.
elliptical galaxies, spiral bulges and halos are all included in the same category, described in this section. 
Here we recall the main assumptions:
ellipticals form by means of a rapid infall of pristine gas, which triggers an intense starburst. 
Star formation  is assumed to halt when  
the energy of the ISM, heated by stellar winds and supernova (SN) explosions,
exceeds the binding energy of the gas (Dekel \& Silk 1986). 
At this time a galactic wind occurs, sweeping away almost all of  
the residual gas present at that time. After the SF has stopped, 
the galactic wind is maintained by Type Ia SNe which continue exploding until the present time, 
and its duration
depends on the balance between this heating source and the gas cooling. 
The binding energy of the gas is strongly 
influenced by assumptions concerning the presence and distribution of dark
matter (Matteucci 1992). For the model adopted here, a diffuse 
($R_e/R_d$=0.1, where
$R_e$ is the effective radius of the galaxy and $R_d$ is the radius 
of the dark matter core) but 
massive ($M_{dark}/M_{lum}=10$) dark halo has 
been assumed (see Bertin et al. 1992). \\

The outflow rate $(\dot{G_{i}})_{out}$  is of the
same order of magnitude (in general within a factor of 2) of
the value taken by the SFR just before the galactic wind (Pipino et al. 2005). 
This result is in agreement with the indications from
the observations of starburst galaxies (e.g. Heckman 2002). For further details, see Pipino et al. (2005).\\
We assume that the efficiency of star formation is higher in more 
massive objects which evolve faster than less massive ones (inverse-wind 
scenario, Matteucci, 1994, otherwise called ``downsizing'').
This implies that the galactic 
wind develops
after a 
timescale varying with the galactic mass, with more massive galaxies experiencing 
earlier outflows. This mechanism allows us to well reproduce the observed increase of [Mg/Fe] with galactic mass in ellipticals (see Matteucci 1994; Pipino \& Matteucci 2004). It is also worth noting that this wind mechanism is the opposite of what proposed originally by Larson (1974) to explain the MZ-relation in ellipticals. In fact, in Larson's model the galactic wind develops later in more massive systems due to an assumed constant efficiency of star formation with galactic mass. Matteucci (1994) obtained the ``inverse-wind'' scenario simply by assuming an increasing efficiency of star formation with mass and such an assumption can preserve the MZ-relation as well. 

A Salpeter (1955) IMF constant in space and time is adopted, with lower limit $0.1 M_{\odot}$ and upper limit $100 M_{\odot}$.

\begin{figure}
\includegraphics[width=0.5\textwidth]{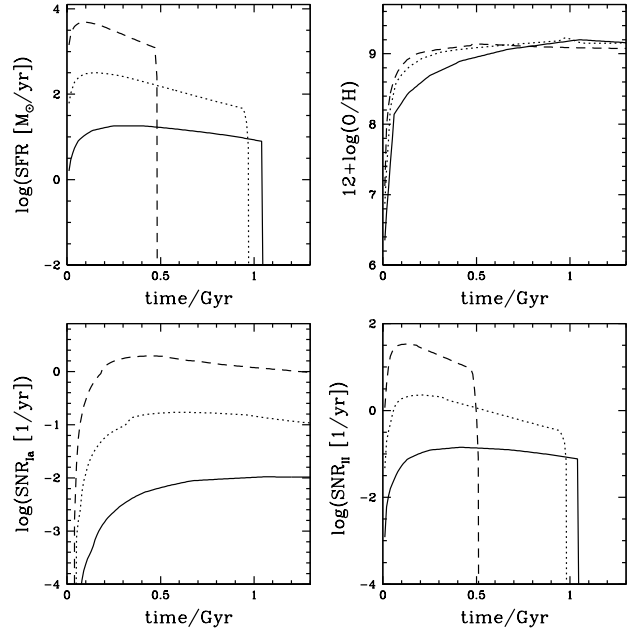}
\caption{From top left corner, clockwise: SFR, interstellar O abundance (in units log(O/H)+12), Type II SNR,
 and Type Ia SNR vs time for the three elliptical galaxy models used in this paper. 
The solid, dotted and dashed lines are the predictions for the models with total baryonic 
mass $10^{10} M_{\odot}$, $10^{11} M_{\odot}$, and $10^{12} M_{\odot}$, respectively. 
\label{SFR_ell}}
\end{figure}

The choice of such an IMF for ellipticals and S0 galaxies assures that 
several observational constraints such as the average stellar abundances and the color-magnitude diagram 
(see Pipino \& Matteucci, 2004) are well reproduced, as well as the metal content in clusters of galaxies (see Renzini 2004; Calura, Matteucci \& Tozzi 2007).  
We consider three elliptical galaxy models of three different luminous masses: 
$M_{lum}=10^{10}, 10^{11}$ and $10^{12} M_{\odot}$.
The infall is assumed to occurr on an extremely short timescale ($< 0.5$ Gyr). 
In the Table 1 we list the initial baryonic mass, 
the effective radius, the star formation efficiency $\nu$, the infall timescale  $\tau_{inf}$ and the IMF adopted for each galaxy.   
In Fig.~\ref{SFR_ell}, we show the predicted time evolution of the star formation rates, Type Ia, Type II SNe rates and interstellar 
metallicity, represented by the O abundance (in units of 12+log(O/H) for the three elliptical galaxy models used in this paper. \\

%
%
%
\begin{table}
\caption{Elliptical models: parameters}         
\label{table1}                                 
\centering                                      
\begin{tabular}{c c c c c }                        
\hline\hline                                    
Baryonic Mass & $R_{eff}$ & $\nu$       &  $\tau_{inf}$       &       IMF     \\         
($M_{\odot}$ ) & (kpc)    & ($Gyr^{-1}$) &   (Gyr)          &               \\
\hline                                          
  $10^{10}$   &   1       &  3          &    0.5                & Salpeter \\                         
  $10^{11}$   &   3       &  10         &    0.4                & Salpeter \\                         
  $10^{12}$   &   10      &  25         &    0.2                & Salpeter \\                         
\hline                                          
\end{tabular}
\end{table}

\subsection{Spiral galaxies} 
To model spiral galaxies, we use a  single-infall 
model designed to describe the evolution of the thin disc of the Milky Way galaxy (Chiappini et al. 2001). 
By using this model, we make the implicit assumption that the baryonic mass of any 
spiral galaxy is dominated by a thin disc of stars and gas 
in analogy with the Milky Way.  
The model used for spiral discs is a multi-zone one.
The disc is approximated by several independent rings, 
2 kpc wide, without exchange of matter between them. 
The timescale for the  
disc formation is assumed to increase 
with the galactocentric distance, thus producing an ``inside-out'' scenario 
for the disc formation (Matteucci \& Francois 1989; Chiappini et al. 2001; Cescutti et al. 2007). 

\begin{figure}
\includegraphics[width=0.5\textwidth]{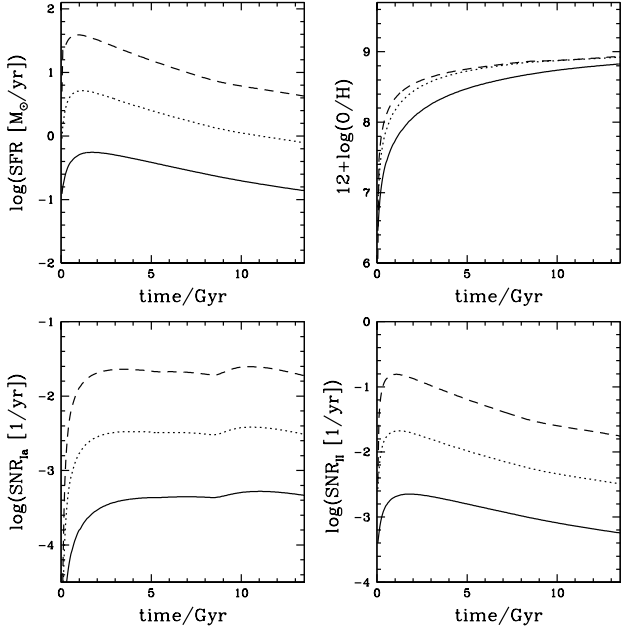}
\caption{From top left corner, clockwise: SFR, interstellar O abundance (in units log(O/H)+12), , Type II SNR,
 and Type Ia SNR vs time for the three spiral galaxy models used in this paper. 
The solid, dotted and dashed lines are the predictions for the models with central surface density
$\Sigma_{0}=200 M_{\odot}/pc^{2}$, $\Sigma_{0}=500 M_{\odot}/pc^{2}$, and  $\Sigma_{0}=2000 M_{\odot}/pc^{2}$, respectively. 
\label{SFR_spi}}
\end{figure}

In this work, we use three spiral models representing galaxies of different masses. 
One model is designed to 
reproduce the main features of the M101 spiral galaxy (see Chiappini et al. 2003), representing 
the most massive spiral disc. 
The MW disc is used to represent  a spiral galaxy of intermediate mass. 
Finally, a model for a low-mass spiral 
has been developed, characterized by a star formation efficiency and a mass surface density 
lower than the ones of the Milky Way (see Tab. 2). \\
Each model is characterized by a particular infall law and a star formation efficiency.
Our main assumption here is that larger discs evolve faster than smaller ones (Boissier et al. 2001), 
in analogy with ellipticals and in agreement with observations.  
For the mass surface density of each spiral disc, we assume an exponential profile 
$\Sigma_{tot}(R)= \Sigma_0 \exp(-R/R_{d})$. The SFR surface density is a 
Kennicutt (1998) law expressed as:
\begin{equation}
\dot{\sigma}_{*}(t) = \nu \sigma_{gas}^{k}(t)
\end{equation}
with $k=1.5$. As for ellipticals, 
the SF efficiency $\nu$ increases with the galaxy mass. 
The IMF is the one of Scalo (1986), and it is assumed to be constant in space and time.
The use of this IMF is motivated by the results by 
by Kroupa \& Weidner (2004), who have shown that, since the disc stellar population is made by dissolving 
open stellar clusters, the disc IMF must be significantly steeper than the cluster IMF, 
because the former results from a folding of the latter with the star-cluster mass function. \\
An important indication about the IMF in spiral discs comes also from chemical evolution models (see Chiappini et al, 1997; 2001), which clearly
indicate that to reproduce the main features of the solar neighbourhood and the whole disc of the Milky Way  
a Scalo-like IMF (constant during the entire disc evolution) is 
preferred, and that the Salpeter IMF would overestimate the solar abundances (see Romano et al. 2005 for a detailed discussion on this point). 
In Table 2 we list the adopted values for the central surface mass density $\Sigma_0$, the scaling radius $R_D$, the star formation efficiency $\nu$, the infall timescale $\tau_{inf}$ and the IMF for spirals of different masses.   
In Fig.~\ref{SFR_spi}, we show the time evolution of the star formation rates, Type Ia, Type II SNe rates and interstellar 
O abundance 
for the three spiral galaxy models used in this paper. 
In spiral discs, to compute the interstellar O abundance by means of our multi-zone models, 
we have calculated the mass-weighted interstellar O abundance on the whole disc:
\begin{equation}
X_{O, ISM}=\frac {\sum_{j} \sigma_{ISM,j} X_{O,j}}{\sum_{j} \sigma_{ISM, j}}
\label{X_O}
\end{equation}
(see Calura \& Matteucci 2004), where the $\sigma_{ISM, j}$ are the ISM surface densities in the various regions of the disc, whereas 
 the $X_{O, j}$
are the gas-phase O abundances in these regions. To compute $X_{O, ISM}$, for each spiral model, 
we have considered only the regions  located at galactocentric distances $3\le R/kpc \le 8$.
This is done since the observational metallicities used in this work have been obtained considering maximum 
apertures of $\sim 9$ kpc, as in Savaglio et al. (2005) at redshift $z\sim 0.7$. 
On the other hand, at the median redshift of 0.07 the SDSS fiber size (3'') 
corresponds to a radius of 3-4 kpc. 
Our assumption of considering only regions at radii $\le 8 kpc$ has a minor impact on our results. In fact, 
it is worth to stress that, 
since the innermost regions
are the densest and most metal enriched ones, 
the interstellar O abundance described by eq.~\ref{X_O} is dominated by the contribution of regions 
within $R \le 4 kpc$. 
We assume that spiral galaxies do not experience galactic winds. 
This assumption is motivated by the results of Tosi et al. (1998), who showed that a
chemical evolution model  of the galactic disc including  mass outflows is not able to 
reproduce the abundances and abundance ratios observed in field stars. 

%
%
%
\begin{table}
\caption{Spiral models: parameters}         
\label{table2}                                 
\centering                                      
\begin{tabular}{c c c c c}                        
\hline\hline                                    
$\Sigma_{0}$          & $R_{D}$   & $\nu$          &$\tau_{inf}(R)$ & IMF     \\         
($M_{\odot}/pc^{2}$ ) & (kpc)    &  ($Gyr^{-1}$)    &    (Gyr)       &           \\
\hline                                          
  200 & 2   & 0.3 &      $R \cdot 1.03-1.27$                          &Scalo \\                         
  491 & 3.5 & 1   &      $R \cdot 1.03-1.27$                          & Scalo \\                         
  2000 & 5 & 2    &      $R \cdot 0.75-0.5$                              & Scalo \\                         
\hline                                          
\end{tabular}
\end{table}

\subsection{Irregular galaxies}
\begin{figure}
\includegraphics[width=0.5\textwidth]{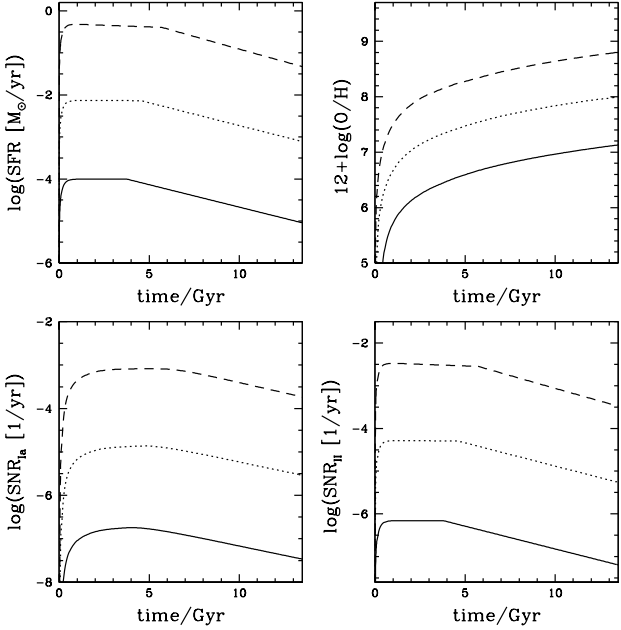}
\caption{From top left corner, clockwise: SFR, O abundance the gas (in units log(O/H)+12), Type II SNR,
 and Type Ia SNR vs time for the three irregular galaxy models used in this paper. 
The solid, dotted and dashed lines are the predictions for the models with baryonic mass 
$10^{8} M_{\odot}$, $10^{9} M_{\odot}$, and $10^{11} M_{\odot}$, respectively. 
\label{SFR_irr}}
\end{figure}

For the irregulars we assume a
one-zone model with instantaneous and complete mixing of gas inside
this zone. 
Irregular galaxies 
assemble all their mass by means of a continuous infall 
of pristine gas.
The SFR is continuous, with SF efficiency values lower than the ones 
used to describe ellipticals and spirals. 
In fact, as suggested by Calura \& Matteucci (2006a), the main features of local galaxies of different morphological types 
may be reproduced by interpreting the Hubble sequence as a sequence of decreasing SF efficiency from early types to late types, 
i.e. from ellipticals to irregulars.\\
Also in this case, the star formation 
can trigger a galactic wind if the thermal energy of the gas exceeds its binding energy (Bradamante et al. 1998, Recchi et al. 2002). 
As for Elliptical galaxies, for Irr the adopted IMF is the one of Salpeter (1955). 
This choice is in agreement with the results by Calura \& Matteucci (2004), who have shown that the 
adoption of a steeper IMF in dwarf irregulars leads to an underestimation of their average metallicity. \\
The rate of gas loss via galactic winds for each element {\it i} 
is assumed to be proportional to the star formation rate at the 
time {\it t}:

\begin{equation}
  (\dot{G_{i}})_{out} =\,w_{i} \, \psi(t)
\label{wind_eq}
\end{equation}

\noindent
where $w_{i}$ is a 
free parameter which describes the efficiency of the galactic
wind for a given element $i$, and it is the same for all the elements.\\
In irregular galaxies, the parameter $w_{i}$, is fixed in order to reproduce the present-day gas fraction observed in dwarf irregulars. 
The assumption of  $w_{i}=0.25$ allows us to account for the correct fraction of neutral gas in irregulars (Calura \& Matteucci 2006a). \\
Also in this case, we model irregular galaxies of three different luminous masses: 
$M_{lum}=10^{8}, 10^{9}$ and $10^{10} M_{\odot}$.  
These values should  bracket the baryonic masses of the most massive typical dwarf Irregular galaxies (Lee et al.  2006), and 
are compatible with the lowest stellar masses of the observational data samples used in this work. \\
Table 3 shows our adopted values for the initial baryonic mass, the luminous radius, the star formation efficiency $\nu$,
the infall timescale and the IMF for each galaxy. For all irregulars, we assume a luminous radius of 1 kpc.    
In Fig.~\ref{SFR_irr}, we show the time evolution of the star formation rates, Type Ia, Type II SNe rates and interstellar 
O abundance for the three dwarf galaxy models used in this paper. \\

%
%
%
\begin{table}
\caption{Irregular models: parameters}         
\label{table3}                                 
\centering                                      
\begin{tabular}{c c c c c c}                        
\hline\hline                                    
Baryonic Mass & $R_{lum}$ & $\nu$       &  $\tau_{inf}$  & $w_i$  & IMF     \\         
($M_{\odot}$ ) & (kpc)    & ($Gyr^{-1}$) &   (Gyr)       &       &          \\
\hline                                          
   $10^{8}$ & 1 & 0.001 &  0.2 & 0.25   &Salpeter\\                         
  $10^{9}$  & 1 & 0.007 &  0.2 & 0.25   &Salpeter\\                         
  $10^{10}$ & 1 & 0.05  &  0.2 & 0.25   &Salpeter\\                         
\hline                                          
\end{tabular}
\end{table}


\begin{figure}
\includegraphics[width=0.4\textwidth]{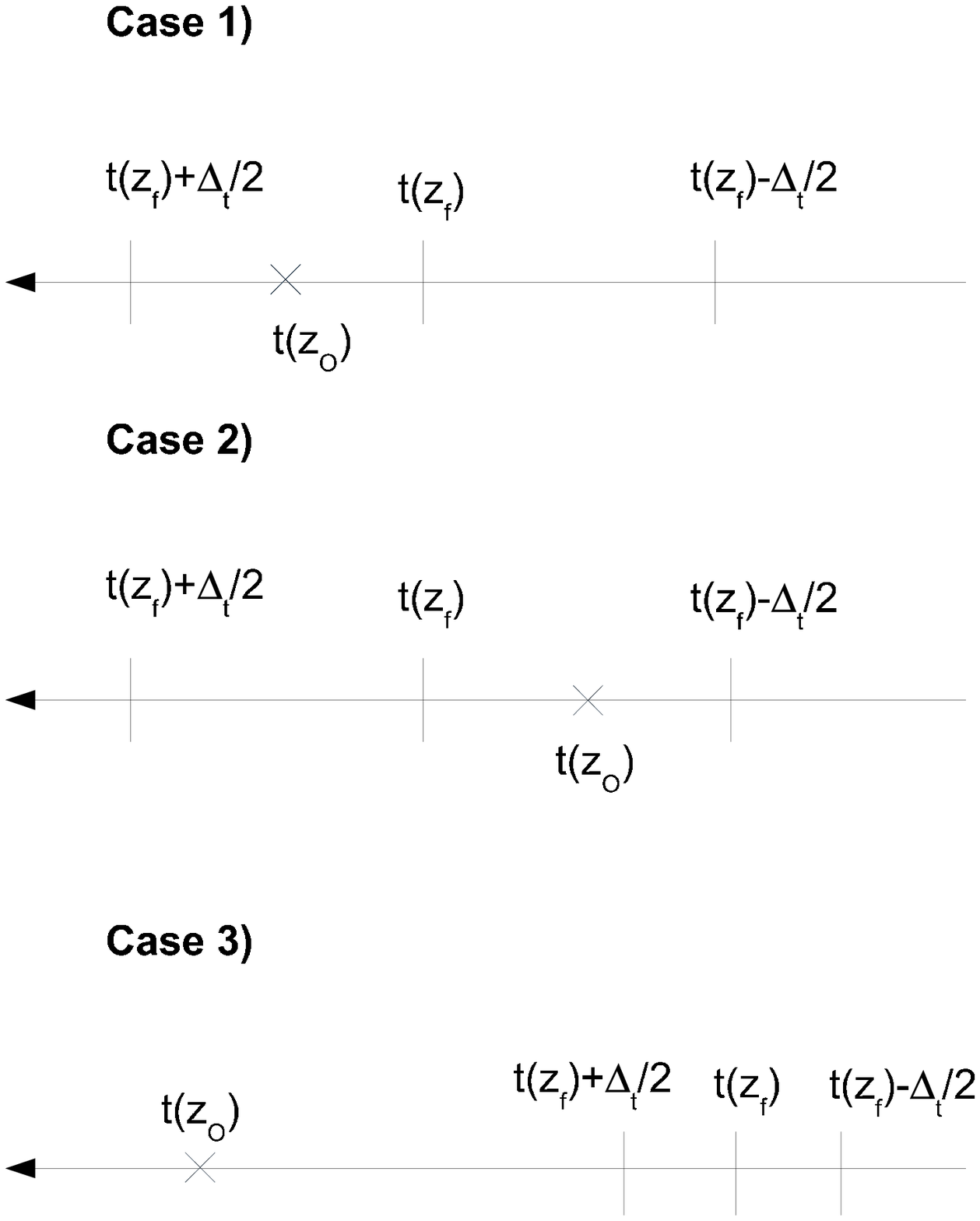}
\caption{In cases 1 and 2, galaxies are still forming at the redshift of observation, i.e. 
$t(z_f)-\Delta_t \le t(z_o) \le t(z_f)+\Delta_t$, where $t(z)$ is the age of the Universe at the redshift $z$. 
In case 3, galaxies have stopped forming at the redshift $z_o$, 
so $t(z_o)>t(z_f)+\Delta_t $. 
\label{tz}}
\end{figure}

\subsection{Different galaxy formation epochs}
\label{random}

As we have just seen, for each galactic morphological type we adopt a set of three models 
correponding to fixed  initial baryonic masses. 
These models are designed to reproduce the main chemical features 
of galaxies of different morphological types. 
In general, in chemical evolution studies, the galaxies are all considered coeval 
in the sense that, to compare any prediction 
with local observations, in general one assumes that ellipticals, spirals and irregulars 
have today an age comparable to the one of the Universe, i.e. $\sim 13.5$ Gyr.
This is justified by the observational fact that the majority of galaxies contains stars as old as the Universe. This is certainly true for the ellipticals and spirals, although spiral discs are generally younger than stellar halos and have ages in the range 8-11 Gyr,  whereas for the irregulars this is less clear. Then, one  considers the model outputs at this age, 
which are compared with observations at $z=0$. \\

However, in a more realistic picture, 
one should allow for some differences among the times at which star formation started, 
in order to take into account a possible intrinsic spread in the MZ-relation. 
For elliptical galaxies, a realistic age formation spread is 3 Gyr as indicated by studies of field ellipticals (Bernardi et al. 1998), 
with the cluster ellipticals showing even a smaller spread of 2 Gyr (Bower et al. 1992). 
For the spirals the situation is less clear: we tentatively assume a spread of 5 Gyr on the basis of the fact that 
discs take a longer time to assemble by gas accretion than spheroids. As we will see later (Sec.~\ref{MZ_SFR}), such an age dispersion values are 
compatible with other age estimates for spirals present in the literature. \\
Concerning irregulars, in principle, the 
age spread could be as long as  $\sim$ 10 Gyr, i.e. comparable to a Hubble time. \\
In this section, we describe how we simulate a population of non-coeval galaxies 
and how we create a finer mass grid. 
First of all, we define as galaxy formation epoch the time at which star formation starts in a given galaxy.
For each morphological type, we choose a mean 
redshift of formation $z_f$ and an age dispersion $\Delta_t$, relative to the spread in the galaxy formation time. 
If the mean redshift of formation is $z_f$ 
and $t(z)$ is the age of the Universe at a given redshift $z$,  
galaxies are allowed to form across 
the cosmic time interval $t(z_f)-\Delta_t/2 \le t \le t(z_f)+\Delta_t/2$, 
i.e. this means that galaxies may start forming at a redshift greater than $z_f$. 
If $z_o$ is the redshift of observation of a particular galactic observable, 
the ages (defined as the times elapsed since the beginning of star formation) 
of the galaxies at $z_o$ are allowed to vary between two values $T_{min}$ and $T_{max}$, depending on $z_o$ and 
the assumed values of $z_f$ and $\Delta_t$: 
\begin{itemize}
\item If $t(z_f)-\Delta_t/2 \le t(z_o) \le t(z_f)+\Delta_t/2$, i.e.   if galaxies are still forming at the redshift of observation, 
$T_{min}=0$ and $T_{max}=abs(t(z_f)-(t(z_{o})+\Delta_{t}/2)$ (Cases 1 and 2 in Fig~\ref{tz})
\item If $t(z_o)>t(z_f)+\Delta_t/2 $, i.e. galaxies have stopped forming at the redshift $z_o$, then 
$T_{min}=t(z_o)-(t(z_f)+\Delta_t/2)$ and $T_{max}=t(z_o)-(t(z_f)-\Delta_t/2)$ (Case 3 in Fig~\ref{tz})
\end{itemize}
where $t(z_f)$ and 
$t(z_{o})$ are the ages of the Universe at the formation redshift  $z_f$  and at the observation redshift $z_o$, respectively.  
For each morphological type and for any given mass, 
we extract random ages from a flat probability distribution 
between $T_{min}$ and $T_{max}$ and we compute
the relevant physical quantities at this randomly extracted age.
With our method, we  can 
simulate a continuous galaxy formation process in  a given time interval and a finite dispersion 
of the galaxy formation epoch. 
At any redshift, 
this method is useful to generate an almost-continuous galactic stellar mass grid.

\begin{figure*}
\epsfig{file=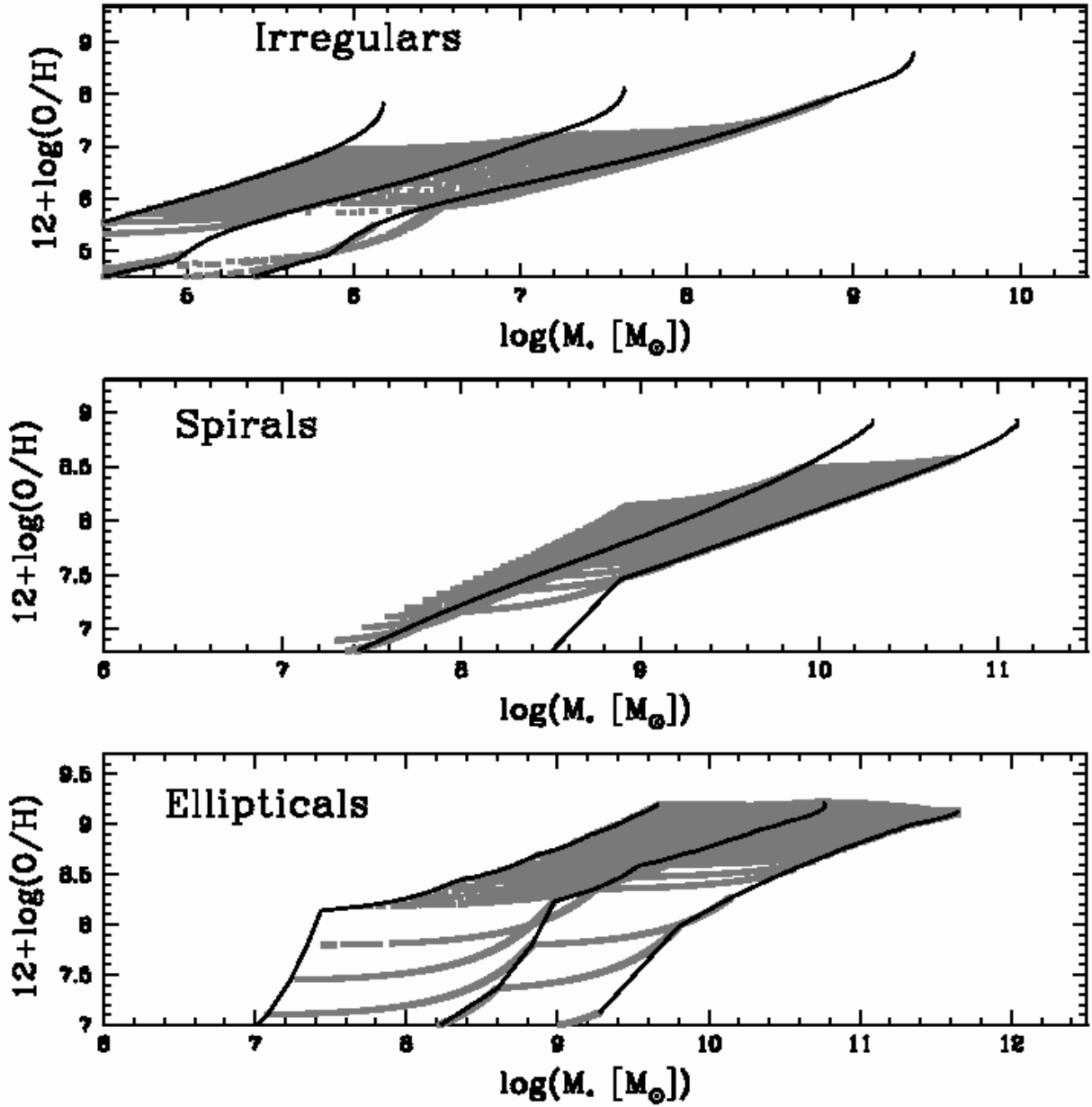, height=18cm,width=18cm}
\caption[]{Predicted MZ relation for galaxies of various morphological types. 
In each panel, the curves represent the evolutionary 
tracks of individual galaxies of various baryonic masses. The grey areas are the 
predictions obtained by means 
of our method to create an almost-continuous galaxy formation and a finer stellar mass grid (see Sect.~\ref{random}). In this case, we have assumed 
$z_{f}=3$, $\Delta_t=3$ Gyr for all galaxies. The predicted MZ is computed at redshft $z_o=2.2$:This choice is due to the fact that at this cosmic epoch also the ellipticals are still star forming galaxies.  
 \label{MZ_exa}}
\end{figure*}

In Fig.~\ref{MZ_exa}, we show an example of the application of this technique: 
the predicted MZ relation for ellipticals, spirals and irregulars, obtained 
assuming $z_{f}=3$, $\Delta_t=3$ Gyr and computed at a redshft $z_o=2.2$.  
For each morphological type, the black curves represent the evolutionary tracks of the baseline models of three baryonic masses, 
whereas the grey regions represent the MZ relation computed by extracting randomly 
stellar masses and O abundances at times between $T_{min}=0$  Gyr and $T_{max}=2.3$ Gyr.  
The discontinuity predicted for elliptical galaxies (lower panel of Fig.~\ref{MZ_exa}) is due to 
the steep increase of the metallicity in the first evolutionary phases, i.e. within the first 0.1 Gyr 
(see Fig.~\ref{SFR_ell}). Also the evolution of the stellar mass density is fast, hence small variation 
in the extracted times correspond to large variations for both the mass and the metallicity. \\
The method described in this section will be used to 
generate mass-metallicity plots for galaxies of various morphological types, to be compared with observational 
data obtained at various redshifts.

\begin{figure*}[h]
\epsfig{file=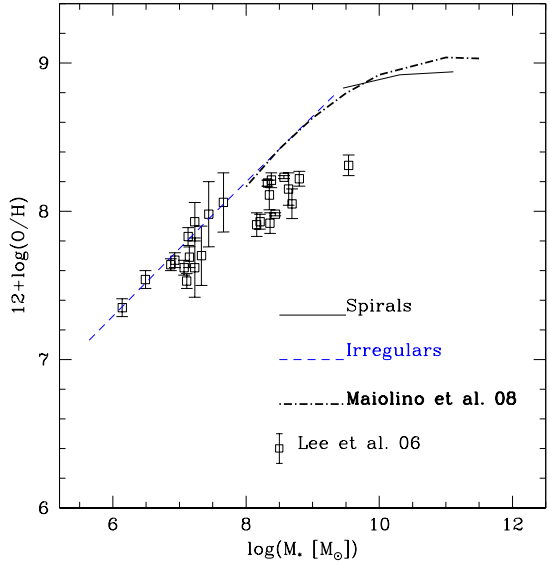, height=8cm,width=8cm}
\caption[]{Predicted MZ-relation at z=0 for spirals and irregulars. 
In the figure is shown also the best fit to the local MZ-relation derived by Maiolino et al. (2008) from the data of Kewley \& Ellison (2008).
The open squares are  the data by Lee et al. (2006), who measured the MZ relation in a sample of local dwarf galaxies. 
\label{MZ_simple}}
\end{figure*}

\begin{figure*}
\centering
\epsfig{file=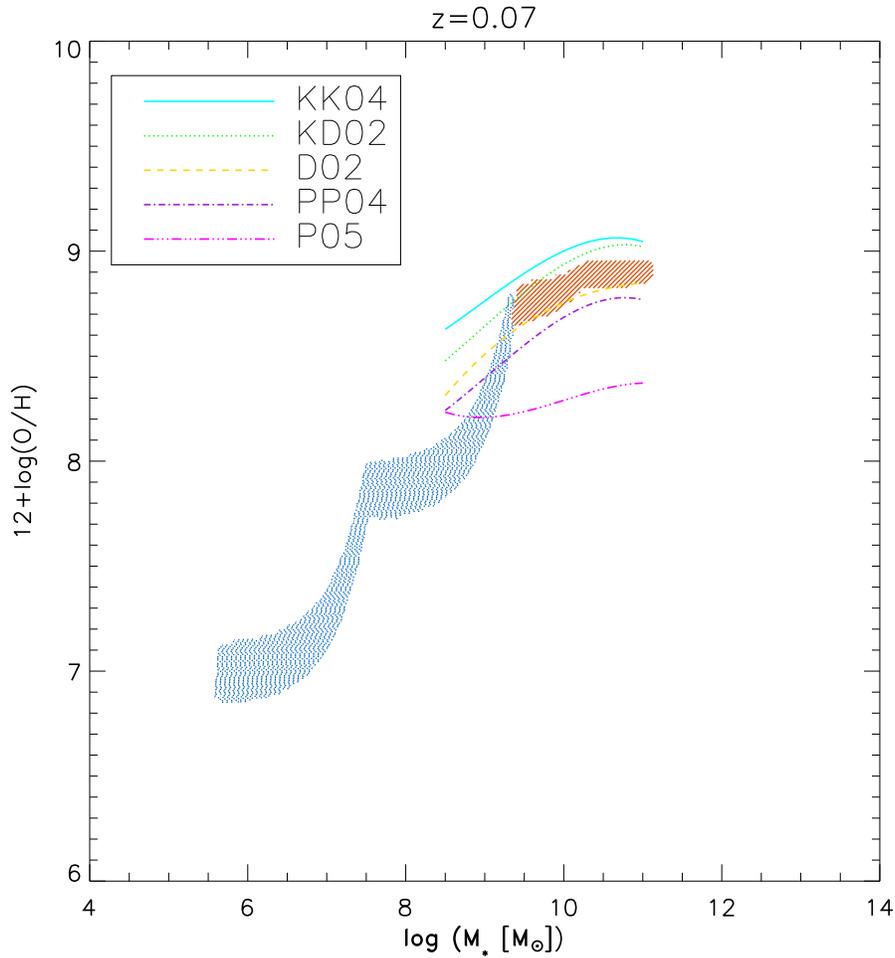, height=13cm,width=13cm}
\caption[]{Predicted Mass-Metallicity for spiral galaxies (red contours) and irregular galaxies (blue contours) at $z=0.07$. 
The solid lines of different colours are best-fit relations calculated by Kewley \& Ellison (2008) using different 
metallicity calibrations. KK04: Kobulnicky \& Kewley (2004); KD02:  Kewley \& Dopita (2002); D02: Denicol\'o (2002); 
PP04: Pettini \& Pagel (2004); P05: Pilyugin \& Thuan (2005). \label{MZ_z0} }
\end{figure*}

\begin{figure*}
\centering
\epsfig{file=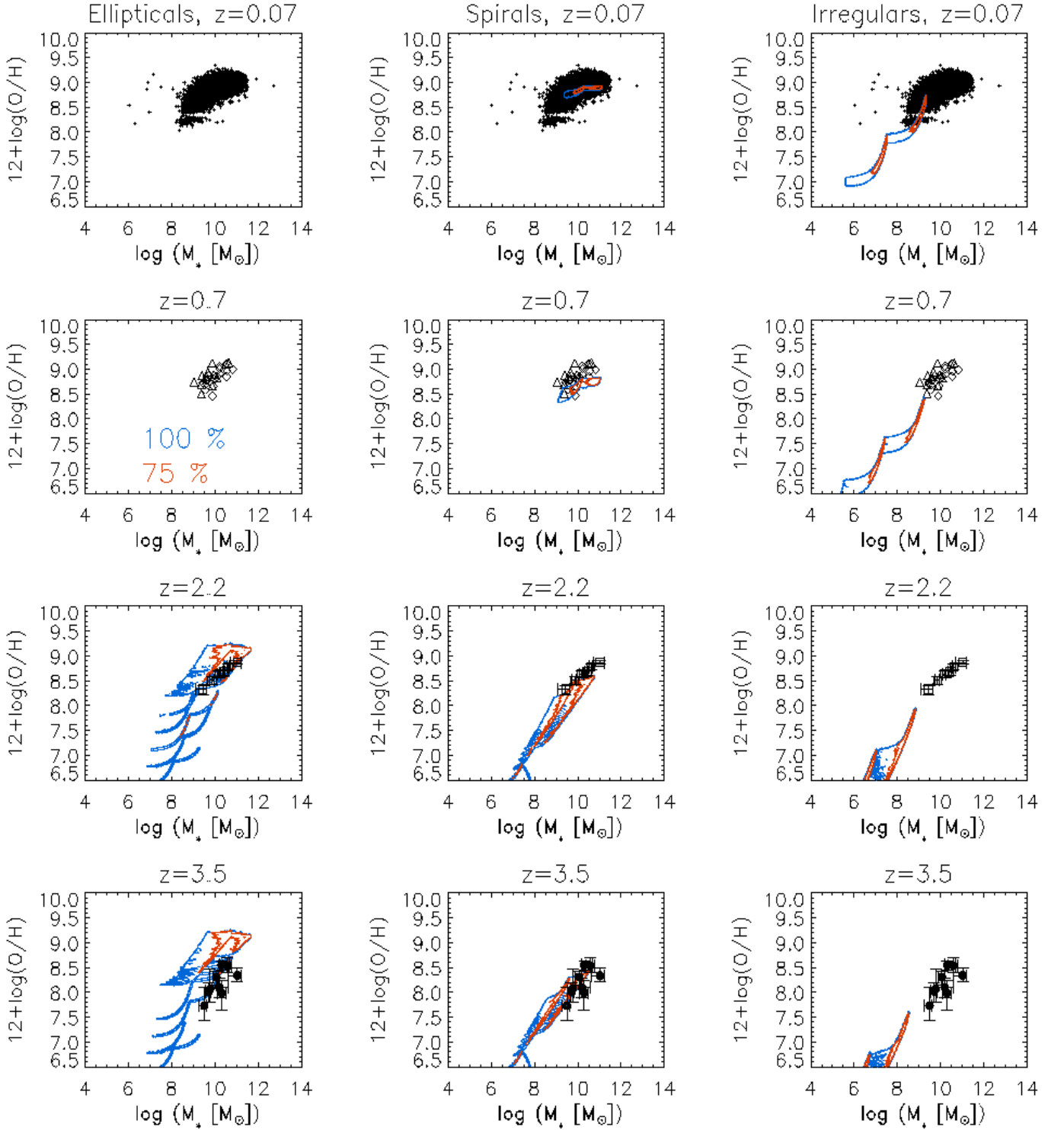, height=20cm,width=19cm}
\caption[]{Redshift evolution of the MZ relation as a function of morphological type. The black points are the observations 
at various redshifts. The blue and red contours repesent the regions where 100 \% and 75 \% of our predictions lie, respectively. 
In the first, second and third column 
the results for ellipticals, spirals and irregulars are shown, respectively. 
From top to bottom and for each morphological type, 
predictions and observations at four different redshifts are shown: $z=0.07$, $z=0.7$, $z=2.2$, $z=3.5$. 
In this case, we have assumed a redshift of formation $z_f=3$ and an age dispersion of $\Delta_t=3$ Gyr for all galaxies. 
Small crosses  at $z=0.07$ are from Kewley \& Ellison (2008);
diamonds and open  triangles at $z=0.7$ from Savaglio et al. (2005); 
the open squares at $z=2.2$ are from Erb et al. (2006a);  
at $z=3.5$, the solid circles are from Maiolino et al. (2008). All observational data have been
recalibrated as discussed in Maiolino et al. (2008).
\label{Mz_morph_1} }
\end{figure*}


\begin{figure*}
\centering
\epsfig{file=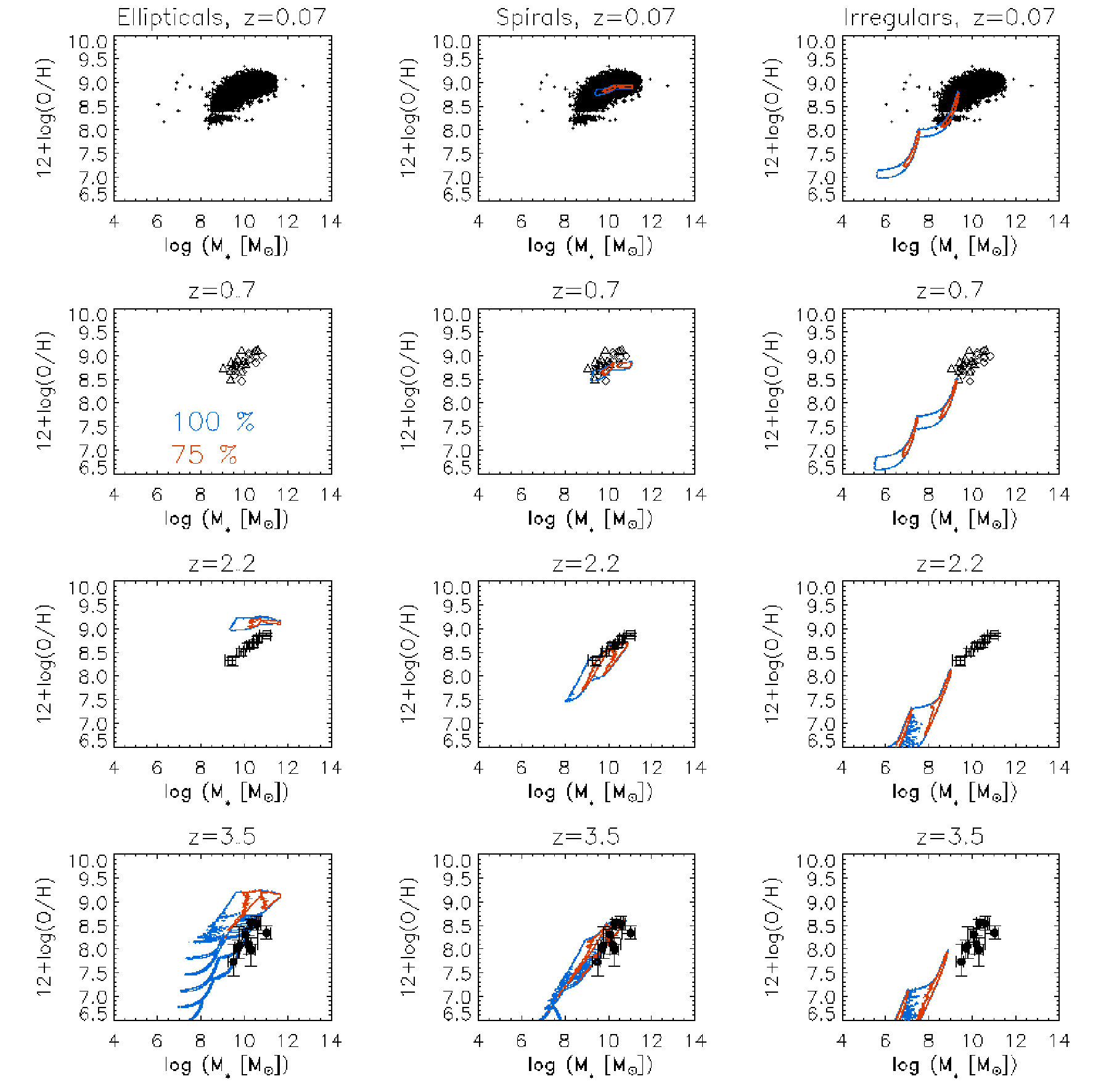, height=20cm,width=19cm}
\caption[]{Redshift evolution of the MZ relation as a function of morphological type. Symbols and contours as in Fig.~\ref{Mz_morph_1}. 
In this case, we have assumed a redshift of formation $z_f=6$ and an age dispersion of $\Delta_t=3$ Gyr. 
\label{Mz_morph_3} }
\end{figure*}

\section{Results}
In Fig.~\ref{MZ_simple}, 
we show the predicted MZ-relations at the present time for spirals and irregulars and compare 
them with the best fit of the MZ-relation obtained by Maiolino et al. (2008) 
from an analithical fit to the local observations (Kewley \& Ellison 2008), 
and to the data of Lee et al. (2006), who determined the MZ relation in a sample of local dwarf galaxies. 
The agreement between our predictions and the observations 
is quite good. 
It is worth to stress that we do not show the results for local ellipticals since  
in this paper we are comparing gas abundances at various redshifts in star forming galaxies, 
so we cannot compare the predicted present time MZ-relation for the gas in ellipticals with the data, 
since these galaxies stopped forming stars several Gyrs ago. 
On the other hand, local ellipticals show a definite MZ-relation,  based on the abundances of their stellar populations, 
which is well fitted by our models (see Pipino \& Matteucci 2004).\\ 
Figure~\ref{MZ_simple} shows that the local MZ relation can be naturally explained by assuming a lower star formation efficiency for less massive galaxies, 
irrespective of the galactic morphological type, and that  galactic winds are not necessary to explain it. 
In fact, our models for spirals do not include galactic winds. 
The reason for this choice is that galactic winds are not required to explain the main features of disc galaxies, as shown by Tosi et al. (1998). 
We consider galactic winds in dwarf irregulars, since they are observed in these systems. 
However, these winds do not carry away large amounts of matter, 
in agreement with the fact that irregulars are gas-rich systems.\\
It is worth to note that 
Dalcanton (2007) reinterpreted the observed MZ-relation in terms of infall/outflow in galaxies 
by means of simple models with IRA and concluded that only gas rich systems with low star formation 
rates can produce and maintain low effective yields. 
This result shows the importance of assuming a lower star formation efficiency in lower mass systems. \\
The flattening of the MZ relation observed at $M_{*}=10^{10} M_{\odot}$ is accounted for by our models and is 
due to the assumed SF efficiencies of the high mass spiral disc models. 
The adopted SF efficiencies cause a 
steep evolution of the (O/H) vs time relation (see Fig.~\ref{SFR_spi}). 
Note that for larger galaxies, their present-day metallicity is reached at earlier times . 
In fact, for the M101 model, very little evolution in the 
(O/H)- time diagram is predicted at cosmic times greater than $\sim 5$ Gyr. The growth of the metallicity  for the MW model is slightly slower than the one 
predicted for M101. This is due to the fact that the adopted SF efficiencies are comparable. 
On the other hand, the model for small discs has a SF efficiency lower by almost one order of magnitude, 
hence the growth of metallicity with time is slower, and the present-day metallicity is much smaller than the ones of the MW and of M101.\\
In conclusion, the observed flattening of the MZ relation for large discs indicates that in the past, their SF efficiences must have been similar. 
This is consistent also  with the fact that the O gradients observed 
in the MW and in M101 are very similar, although M101 is more massive (see Chiappini et al. 2003).

\subsection{The calibration of the mass-metallicity at redshift zero} 
\label{calib}
In Fig.~\ref{MZ_z0} 
we show the predicted MZ relation for local star forming galaxies, compared with 
a set of observational MZ relations at $z=0.07$.   
In our models,  only spirals and irregulars experience 
star formation at $z=0.07$. This is supported by several observational studies, 
which  have shown that local ellipticals are dominated by old stellar populations 
(de Freitas Pacheco,  Michard, \& Mohayaee 2003 and references therein). 
The filled contours  in Fig.~\ref{MZ_z0} are the predicted 
MZ relation computed at z=0.07 for spirals and irregulars, 
whereas the lines are fits to the observed local 
mass-metallicity relation as reported by Kewley \& Ellison (2008), 
obtained assuming different metallicity calibrations. \\
Kewley \& Ellison considered $\sim 28000$ star-forming galaxies of the SDSS 
with derived stellar masses. 
They applied 10 different metallicity calibrations to the emission lines of the spectra of this sample, 
in order to investigate 
the effect of the calibration on the MZ relation. Some calibrations are based on a direct method for determining 
metallicity. This method, often referred to as the "$T_{e}$" method, 
is based on the measure of the ratio of two lines, such as 
the O4363 auroral line and the [OIII]$\lambda$ 5007 lines and allows an estimate of the electron temperature of 
the interstellar gas, finally used to determine the metallicity.
Some other calibrations are empirical, and are obtained by fitting the relation between direct $T_{e}$ metallicities and strong 
line ratios for $H_{II}$ regions, such as [NII]$\lambda$6584/H$\alpha$. 
Finally, in other cases, mostly at high metallicity, where $T_{e}$ is not measurable, 
photoionization models are used to compute strong-line ratios. 
Is it important to stress that each of these calibrations are valid in relatively narrow metallicity intervals (Kewley \& Ellison 2008). \\
In Fig.~\ref{MZ_z0}, the curves labeled KK04 (Kobulnicky \& Kewley 2004), KD02 (Kewley \& Dopita 2002) are based on theoretical methods, 
the curves labeled PP04 (Pettini \& Pagel 2004) and P05 (Pilyugin \& Thuan 2005)) 
are based on empirical methods, while the D02 (Denicolo, Terlevich \& Terlevich 2002) 
curve has been derived by means of a combined method. The sample of D02 
is composed by a set of $H_{II}$ regions, some of which 
have metallicities derived using the $T_e$ method, while some others have 
empirical or theoretical metallicities. \\
As can be seen in Fig.~\ref{MZ_z0}, the choice of the metallicity calibration 
plays an important role in determining observationally 
the MZ relation. Kewley \& Ellison (2008) have outlined that different calibrations 
can produce very different results, concerning both 
the zero-point and the slope of the observational MZ relation, as well as the location of the plateau for high-mass galaxies.  
For the particular cases reported in Fig.~\ref{MZ_z0}, including also the most extreme calibrations,  
one can see that the zero point of the observed MZ relation can vary by $\sim 0.4$ dex, whereas 
the high-mass plateau position may vary  by  $\sim 0.3$  dex, without considering the extreme calibration 
by Pilyugin \& Thuan (2005), leading to an almost flat MZ. \\ 
Concerning the predictions, here we have assumed  $z_f=3$ and $\Delta_t=5$ Gyr for spirals, and 
 $z_f=3$ and $\Delta_t=10$ Gyr for irregulars. 
However, as we will see later, for spirals and irregulars the parameters $z_f$ and $\Delta_t$  
have little effect on the zero point and on the slope of the predicted 
mass-metallicity relation, whereas they mostly influence the predicted dispersion of the stellar masses and of the O abundances. 
As discussed in Sect. ~\ref{models}, 
the models used for spirals and irregulars are calibrated in order to reproduce a large set of 
chemical evolution constraints from local observations, such as the abundances observed in stars 
of the Milky Way, the stellar metallicity distribution in the solar neighbourhood (Chiappini et al. 2001), 
the abundance gradients in the MW and in M101 (Chiappini et al. 2003), 
as well as the stellar abundances and stellar metallicity distributions 
in local dwarf galaxies (Lanfranchi \& Matteucci 2003). Since our models can reproduce such 
a large amount of local independent observations, it seems appropriate to use them to find some constraints 
on the best calibration  method for the local observational MZ relation.
Among the different calibrations considered here, our predictions are compatible 
with the MZ relations obtained by adopting the calibrations KD02, D02 and PP04. \\
The possibility to use our models to constrain the metallicity calibrations is motivated by the fact that most of the observables 
cited above and used to tune our models have typical errors much lower than the calibration uncertainty.  
For instance, the abundances observed in local stars have typical errors of 0.05-0.1 dex, in extremely rare cases larger than 0.2 dex 
(Cayrel et al. 2004; Spite et al. 2005). Similar uncertainties are typical also for abundances observed in the stars of local dwarf 
spheroidal galaxies (Shetrone et al. 2003) and blue compact galaxies (Izotov \& Thuan 1999) and in local early type galaxies 
(Thomas et al. 2005). \\
Concerning the calibration method used by Lee et al. (2006) for the data reported in Fig.~\ref{MZ_simple}, for most 
of the objects the $T_e$ method was used, consistent with the one of Maiolino et al. (2008) for the metallicity 
range investigated by Lee et al. (2006). For a very limited number of systems, Lee et al. (2006) used the strong line method. 
Given the uncertainties of various calibration methods in particular  at metallicities $log(O/H)+12\le 8.2$, 
we neglect this difference and use the data by Lee et al. (2006) at face value.\\

\subsection{The observational data used in this work}
\label{Data}
The observational data for the MZ relation at redshits z=0.07,
z=0.7, z=2.2 and z=3.5 are taken from Kewley \& Ellison (2008),
Savaglio et al. (2005), Erb et al. (2006a) and Maiolino et al. (2008),
respectively. These are all MZ relations for the gas phase metallicity,
and include only star forming galaxies (active star formation is required
to ionize the gas which produce the nebular lines used for the metallicity
determination). In these works different diagnostics and
calibrations have been adopted, which affect the zero point and slope
of the MZ relation. This issue is discussed extensively by
Kewley \& Ellison (2008), who also propose a criterion to interlace
different calibrations. However, each of these calibration is appropriate
in different, relatively narrow metallicity ranges.
Maiolino et al. (2008) note that the metallicy range spanned by galaxies
through the cosmic epochs is so wide (7.7$<$12+loh(O/H)$<$9.1) that no single
metallicity calibration method is appropriate to cover it. As a consequence,
Maiolino et al. (2008) define a new calibration that adopts
the electron temperature method at low metallicities (12+log(O/H)$<$8.3)
and the photionization models by Kewley \& Dopita (2002) at high
metallicities (12+log(O/H)$<$8.3). Details on this method and
the appropriateness of this calibration scale are given in Maiolino et al.
(2008). To have the metallicities of all galaxies at the various redshifts
sampled by the different studies on a consistent metallicity scale,
Maiolino et al. (2008) re-determined the
metallicities of the previous works at lower redshifts by using the new
calibrations inferred by them. Maiolino et al. (2008) also correct the
stellar masses of different studies to the same IMF.
In the following section, we will use the same data as the ones 
used by Maiolino et al. (2008), together with SFR determinations at various redshifts. 
As we will see later, the combined use of observational mass-metallicity measured and of SFR determinations 
will allow us to put some constraints on the morphologies of the galaxies populating the observed MZ relation
at various redshifts.

\begin{figure*}
\centering
\epsfig{file=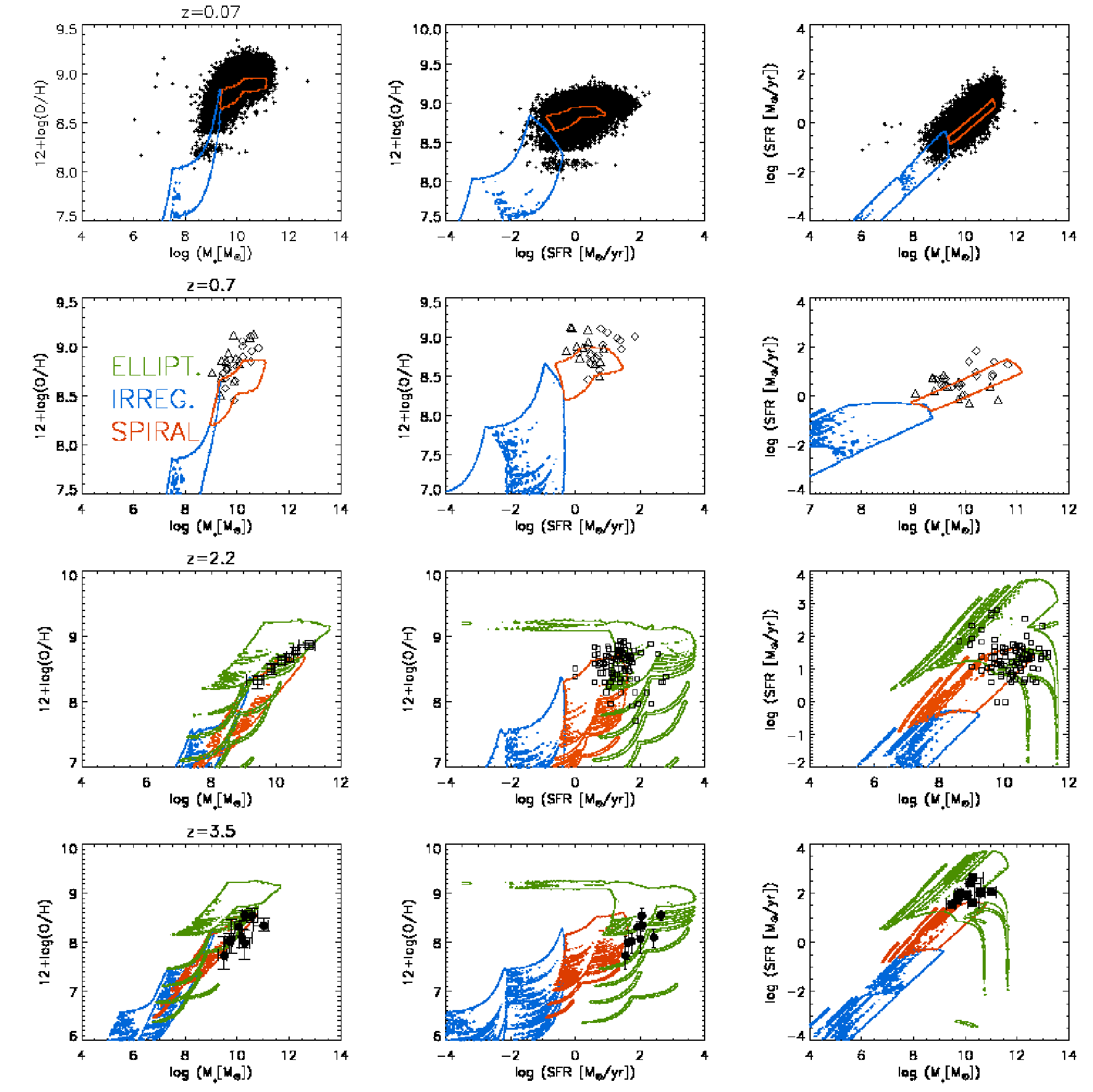, height=23cm,width=19cm}
\caption[]{Predicted redshift evolution of the MZ, (O/H) vs SFR and SFR vs mass plots for ellipticals (green contours), 
spirals (red contours) and irregulars (blue contours)  
and as observed by various authors at $z=0.07$, $z=0.7$, $z=2.2$, $z=3.5$.  
For the predictions, we have assumed a redshift of formation $z_f=3$ and an age dispersion of $\Delta_t=3$ Gyr for 
ellipticals,  $\Delta_t=5$ Gyr for spirals and  $\Delta_t=10$ Gyr for  irregulars. 
Crosses  at $z=0.07$ from Kewley \& Ellison (2008) for the M-Z plot, Kewley \& Ellison (2008) and 
Brinchmann et al. (2004) for the (O/H) vs SFR plot, and Brinchmann et al. (2004) for the SFR vs mass plot. 
Diamonds and open  triangles at $z=0.7$ from Savaglio et al. (2005) and Maier et al. (2005). At this redshift, 
the SFRs are from Juneau et al. (2005). 
At $z=2.2$, the open squares are from Erb et al. (2006a) and Erb et al. (2006b). 
At $z=3.5$, the solid circles are from Maiolino et al. (2008). 
\label{all} }
\end{figure*}

\subsection{The evolution of the mass-Metallicity relation}
\label{evol}
In Figure~\ref{Mz_morph_1} we show   
the observed redshift evolution of the MZ relation in star forming 
galaxies, as well as the predicted MZ relation for various 
galactic morphological types used in this work. 
The predicted MZ relations for ellipticals, spirals and irregulars have 
been computed by assuming $zf=3$, $\Delta_t=$3 Gyr. 
The black points are the observations 
at various redshifts, described in Sect.~\ref{Data}, whereas the blue and red contours represent the regions where 100$\%$ and 75 \% of 
our predictions lie, respectively. 

From this plot, it is clear how galaxies of various morphological types occupy different regions of the MZ plot. 
At any redshift, in general, elliptical galaxies present the highest stellar masses and the highest 
metallicities, whereas the irregulars are the least massive galaxies, 
characterized by the lowest O abundances. In this case, ellipticals appear only 
at $z=3.5$ and at $z=2.2$ since, with the age dispersion of 3 Gyr as chosen here, these galaxies become passive 
at redshift $\le 1.4$. 
From Fig.~\ref{Mz_morph_1}, we can see that the majority of ellipticals show 
metallicities higher than the observations, both at $z=3.5$ and at $z=2.2$. 
This fact concerns only ellipticals and may 
be due to the fact that the growth of (O/H) vs time is very steep for elliptical galaxies. 
We will investigate this 
issue in deeper detail later, in Sect.~\ref{dust}. On the other hand, spiral galaxies show 
metallicities always comparable to the values observed at various redshifts. 
Irregular galaxies have low star formation rates and at high redshift their stellar masses are very small, 
in general $M_{*}\le 10^9 M_{\odot}$ at $z\ge 2.2$. The observed masses of the galaxies populating  
the MZ relation at these redshifts are in general higher than $M_{*}\sim 10^9 M_{\odot}$, hence 
there is poor overlap between the observational MZ relations and our predictions for irregulars at $z\ge 2.2$. 
At $z=0.07$, the SDSS catalogue contains also a lot of galaxies with stellar masses $M_{*}\le 10^9 M_{\odot}$. 
The galaxies with these stellar mass values overlap well with our predictions for irregular galaxies. 
At $z\le 0.7$, we note also that the chosen value for the age dispersion does not allow us to 
reproduce the observed data dispersion. Also this aspect will be investigated in deeper detail later on. \\
A comparison between Fig.~\ref{Mz_morph_1} and Fig.~\ref{Mz_morph_3} is useful to appreciate  
the effects of our parameter 
$z_f$  on our predictions for the MZ relation as a function of redshift. 
The results for spirals and irregulars 
are fairly insensitive on the assumption of the redshift of 
formation $z_f$, whereas for ellipticals the  assumption of an high 
formation redshift, such as $z_f=6$ in Fig.~\ref{Mz_morph_3}, leads to a population 
of extremely high-metallicity ellipticals at $z=2.2$, which does not 
match the MZ relation observed at this redshift by Erb et al. (2006a). \\
It is maybe worth stressing that the contours are not linked to the comoving densities 
of galaxies of various morphological types present at any redshift. 
Such a prediction would require the adoption of  a galaxy luminosity function or a stellar mass function as a function of redshift. 
In this case, the predictions represent the number of galaxies present in each M-Z bin, normalized to the total amount of simulated 
galaxies. 
The blue contours enclose the region where at least 1 galaxy is present. On the other hand, the red contours 
enclose the region where the total number of galaxies is 0.75 times the total number of simulated galaxies. 
In order to relie on a robust statistics, we have simulated $\sim 10^{5}$ galaxies, 
i.e. performed $\sim 10^{5}$ interpolations on the mass and metallicity grids according to the method described in Sect.~\ref{random}. 
However, we have verified that in most cases, the shape of the contour plots is not strongly dependent on the adopted number of simulated galaxies. 
The adopted number of simulated galaxies may have some effects in regions of the plots where galaxies are rare, correspoding to 
the discontinuities visible in the contours. An example is the peculiar ``comb'' feature present in the M-Z plots of elliptical galaxies 
in Fig. ~\ref{Mz_morph_1}, on the left bottom side of the plot. \\
As shown by  Calura et al. (2009), the use of a cosmological galaxy formation 
model can provide directly galaxy abundances in the MZ plot, but no indication concerning the galaxy morphology.  
In the following, we will present a method to constrain the morphology of galaxies, starting from the assumption  that morphology does not change with 
redshift.

\subsection{The mass-metallicity relation of star-forming galaxies: constraints on their morphology}
\label{MZ_SFR}
In this section, we focus on the observed evolution of the mass-metallicity relation for 
star-forming galaxies, considering also their star formation rates. The SFR 
provides us with a fundamental information, since, as we will see, it allows us to have further hints on 
the nature of the galaxies building  the MZ relation. Furthermore, a study of a 3-dimensional 
plot linking mass, metallicity and SFR is a crucial test to our models, which should be able to 
reproduce at the same time all of these properties at any redshift. \\
In Fig.\ref{all}, we show the observed and predicted MZ,  (O/H)-SFR and SFR-Mass plots for all morphological types 
as a function of reshift. 
At each redshift, 
the observed quantities are compared to theoretical 
predictions, obtained for ellipticals, spirals and irregular   
galaxies as explained in sect.~\ref{random}. 
For the observational SFRs, at $z=0.07$ we use the values derived 
for SDSS galaxies by Brinchmann et al. (2004). 
At  $z=0.7$  we use observations by Maier et al. (2005) and 
Juneau et al. (2005), at $z=2.2$ we use the values by Erb et al. (2006b) and at $z=3.5$ we use the data by Maiolino 
et al. (2008). It is important to note that, at $z=0.07$, $z=0.7$ and $z=3.5$ we use metallicities, stellar masses 
and SFRs estimated observationally for individual galaxies. 
In particular, at $z=0.7$, we use 
metallicity measurements only for those galaxies whose SFRs are available. 
At $z=2.2$, the 
measurements of O/H for the individual galaxies are not available. The only data available are represented by 
the rebinned MZ relation as published by Erb et al. (2006a). On the other hand, 
the SFRs for individual 
galaxies are available (Erb et al. 2006b). For these galaxies, we calculate the 
metallicity from the stellar mass, by means of  
the relation found by Maiolino et al. (2008) for star-forming galaxies at $z=2.2$, which 
basically 
represents an analithical fit to the data of Erb et al. (2006a). \\
For the parameters $z_f$ and $\Delta_t$, we have investigated several cases, in the ranges  $2\le z_f \le 6$  and 
 $2$ Gyr $\le \Delta_t \le 10$ Gyr. 
In Fig.~\ref{all}, we show the case computed with the fiducial set of the parameters, i.e.  $z_f=3$ for all galaxies, 
 $\Delta_t=3$ Gyr for elliticals, $\Delta_t=5$ Gyr for spirals and 
$\Delta_t=10$ Gyr for irregulars. It is worth to note that for irregulars and spirals, the assumption of a smaller $\Delta_t$ does not change  
the shape of the MZ relation, of the O/H vs SFR and of the SFR vs $M_{*}$ relations, but affects only the predicted dispersion. 
This plot shows how most of the observational constraints considered in this work 
are satisfactorily reproduced once all morphological types are included.
In this figure, the predictions for ellipticals, spirals and irregulars are shown with different colours. 
At $z=3.5$, the observed MZ, metallicity vs SFR and SFR vs stellar mass plots are well reproduced by our 
models for elliptical galaxies, whereas the spirals and irregulars present SFRs considerably lower than the observations. 
This indicates that the galaxies  
observed by Maiolino et al. (2008) are likely proto-ellipticals observed during the starburst phase. \\
At $z=2.2$, a morphological mix of spirals and ellipticals can reproduce the MZ relation, the (O/H)-SFR and the SFR-Mass plots 
as observed by Erb et al. (2006a, 2006b). 
However, the figure indicates that there is a limited number of observational points which 
are compatible with both predictions for ellipticals and spirals. 
This leads us to a degeneracy concerning the interpretation of the observational results. A possible way to break this degeneracy 
will 
be suggested later in Sect.~\ref{alpha_fe}. \\
At $z=0.7$, by means of the models for spirals we slightly underestimate the metallicities in the MZ and in the 
(O/H) vs SFR plots, however, given the uncertainties due to calibration discussed in Sect.~\ref{calib}, we do not 
consider this as a major issue. 
On the other hand, in the SFR vs mass plot, we predict a positive correlation between 
$M_{*}$  and the SFR. 
The observational data seem to suggest a weaker correlation between $M_{*}$  and the SFR. 
Furthermore, several observed SFRs are higher than the values we predict for spiral galaxies. 
This discrepancy could be partly reduced by considering larger age dispersions for spirals. 
For spirals, assuming a redshift of formation $z_f=3$, 
an age dispersion of $\Delta_t=5$ Gyr implies present  ages 
between 8.86 Gyr  and  one Hubble time. 
Independent estimates of the present ages of spiral galaxies indicate values ranging from 5-6 Gyr up to one Hubble time 
(Bell \& de Jong 2000; Boissier et al. 2001), with low mass spirals younger than high mass discs. 
This  indicates  
that the assumed age dispersion  for spirals may be slightly underestimated, in particular for low mass spirals. 
However, the observed SFRs may be higher owing to episodic 
starbursts, 
possibly triggered by a dynamical process such as galaxy interactions (Alonso-Herrero, Rieke \& Rieke 1998). 
These events, which are of random nature and which are not taken into account by our models, could 
be the cause of the weak correlation between SFR and $M_{*}$ as observed at $z\sim0.7$.
For this reason, this explanation may be the most likely for 
the discrepancy between our predictions and the observed SFRs at $z=0.7$.\\
At $z=0.07$, we can reproduce the shape of the observed MZ relation. However, the dispersion 
of the data is larger than what our predictions indicate. 
For spirals, by assuming $\Delta_t=5$ Gyr, we predict a maximum dispersion of 0.2 dex. 
The observed dispersion cannot be reproduced even assuming the unrealistic value of $\Delta_t =14$ Gyr, i.e. equal to an Hubble time. 
Also in this case, 
this discrepancy 
could be due to the fact that the star formation histories considered here for spirals and irregulars 
do not take into account stochastic events such as 
episodic starburtsts, which could increase the dispersion of the observed metallicities.
However, this does not represent a major concern since, as shown in Kewley \& Ellison (2008), 
also the dispersion depends on the calibration. 
The use of the metallicity calibration of KD02, along with some others (e.g. Pilyugin \& Thuan 2005) 
produce dispersions in the local MZ relation of $\sim 0.6$ dex, whereas with 
other calibrations the degree of dispersion in the data is lower ($\sim 0.4$ dex or lower,Kewey \& Ellison 2008). 
For this reason, in this study, 
the observed y-axis dispersion in the MZ or (O/H) vs SFRs cannot be considered as a key-constraint to our models. \\
Also at $z=0.07$, a large number of observed galaxies present SFRs higher than what our predictions 
for spirals indicate. 
However, the observed correlation between SFR and $M_{*}$ is here reproduced by our predictions. 
Interestingly, at $z=0.07$ the data indicate SFRs even higher than those observed at $z=0.7$. 
Locally, very high SFR may be measured in 
luminous infrared galaxies or in galaxies experiencing episodic starbursts, not considered here. 
However, it is possible that, for some galaxies, the star formation rates may be overestimated. 
In fact, by comparing to other data of star-forming galaxies observed in the AEGIS field (Noeske et al. 2007),
in the lowest redshift bin the maximum observed SFRs are 
$log (SFR/M_{\odot}/yr) \sim 1.3 $, consistent with our values predicted for spirals, 
whereas the SFRs of the SDSS data may be as high as 
$log (SFR/M_{\odot}/yr) \sim 1.8 $ or, in a very few cases, even higher. We suggest here that the discrepancy
 may be due to different dust extinction corrections. This aspect will be investigated in detail with next release of the SDSS data.

\begin{figure*}
\centering
\epsfig{file=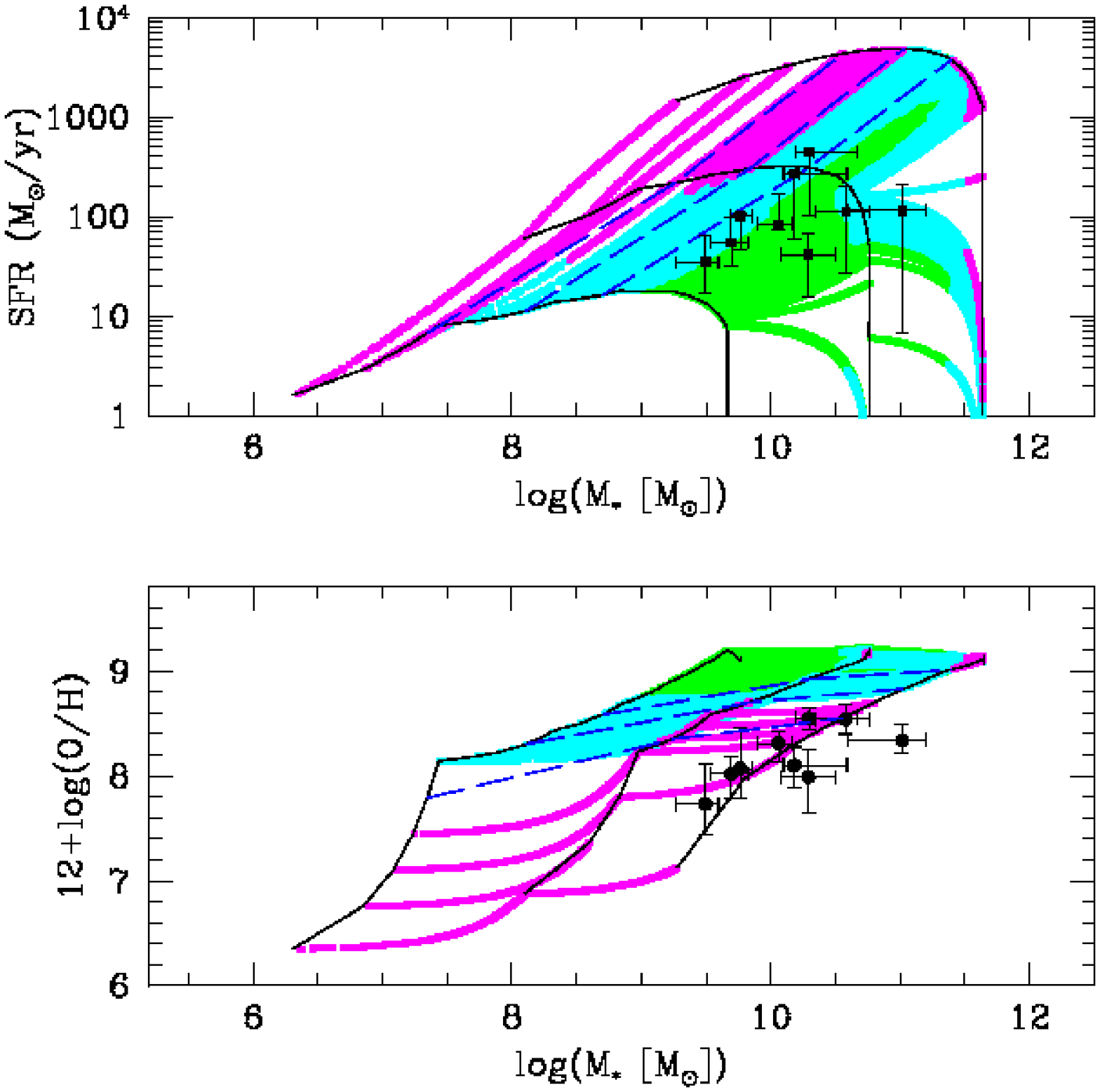, height=16cm,width=16cm}
\caption[]{Upper panel: SFR vs stellar mass plot at $z=3.5$. Green, magenta and cyan regions as above. 
The three dashed lines are isochrones at 0.05 Gyr, 0.1 Gyr, and 0.2 Gyr, from the 
left to the right of the plot. The three solid lines are the evolutionary tracks for the three 
baseline models for ellipticals considered in this paper. \\
Lower panel: MZ plot at $z=3.5$ as predicted for elliptical galaxies and as observed by Maiolino 
et al. (2008). The green, cyan and magenta regions represents our predictions for ellipticals with 
dust surface mass densities in the ranges $\sigma_d> 0.3 M_{\odot}/pc^2$,  $0.1 \le \sigma_d/ 0.3 M_{\odot}/pc^2 \le 0.3$, 
 $\sigma_d< 0.1 M_{\odot}/pc^2$, respectively. 
The three dashed lines are isochrones at 0.05 Gyr, 0.1 Gyr, and 0.2 Gyr, from the 
lowest line to the highest one. The three solid lines are the evolutionary tracks for the three 
baseline models for ellipticals considered in this paper. \\
\label{Mz_dust} }
\end{figure*}

\subsection{A possible dust-obscuration bias at $z=3.5$}
\label{dust}
In Sect.~\ref{evol}, when discussing the MZ relation at $z=3.5$, 
we have seen that our predictions for ellipticals, i.e. the best candidates for the galaxies observed at $z=3.5$ by Maiolino et al. (2008), 
indicate metallicity values higher than the observed ones by $\sim0.5$ dex.  
This difference is higher than the offset due to different calibrations discussed in Sect.~\ref{calib}, since the metallicity 
calibration we use in this paper provides metallicities very close to the ones of Kewley \& Dopita (2002), 
one of the calibrations  providing 
the highest metallicities in the set analyzed by Kewley \& Ellison (2008). 
This fact concerns only the predictions for ellipticals galaxies. 
The discrepancy between the metallicities predicted for most ellipticals at $z=3.5$ and the observed values 
may be partly due to dust obscuration. 
To investigate this issue, we use chemical evolution models for ellipticals 
including also dust grain production and destruction. 
The method to model dust evolution in galaxies is described in 
Calura, Pipino and Matteucci (2008a). 
Dust grains are produced in low mass stars, Type Ia and Type II SNe, destroyed by SN shocks and the grain cores are allowed to accrete mantles, i.e. to grow by mass, during the starburst, when large reservoirs of molecular H is available for star formation. For further details on the formalism built to handle dust evolution in galaxies of different morphological types, we refer the reader to the paper by Calura et al. (2008a).

In this case, we are interested in the dust surface density at various evolutionary phases. 
In ellipticals, the dust surface mass density $\sigma_d$ is 
\begin{equation}
\sigma_d = M_d /\pi R_{eff}^{2}
\end{equation}
where $M_d$ is the total dust mass, including the refractory elements C, O, Fe, S, Si, Fe and Mg 
(Calura et al. 2008a), and $R_{eff}$ is the effective radius. \\

A study of the amount of dust present in proto-ellipticals at various evolutionary stages is interesting because of the link between 
dust mass and obscuration (Calzetti 2001). Stages characterized by higher dust surface densities 
correspond to more dust-obscuration, hence in principle to phases more difficult to observe.  
However, as stressed before, owing to the intense UV radiation field present in $H_{II}$ regions, 
dust depletion is not likely to affect the measured metallicities (Okada et al. 2008).  \\
In Fig.~\ref{Mz_dust}, we show the predictions for the MZ plot and the SFR vs Mass plot 
for ellipticals at $z=3.5$ grouped in three different 
regions, depending on the dust column density at various evolutionary stages. 
The data points with error bars are the observations of Maiolino et al. (2008). 
The three different colours of the models indicate predicted galaxies with $\sigma_d \le 0.1 M_{\odot}/pc^2$ (magenta), 
$0.1 < \sigma_d /M_{\odot}/pc^2\le 0.3$ (cyan), $\sigma_d > 0.3 M_{\odot}/pc^2$ (green). 
The solid lines represent the single evolutionary tracks for the three models 
used to describe elliptical galaxies. The dashed lines are isochrones at times 0.05 Gyr, 
0.1 Gyr and 0.2 Gyr. \\
In the MZ plot, most of the data points overlap with the predictions 
computed for ellipticals at the earlier phases,  
characterized by the lowest dust surface densities. 
The fact that some stellar masses are underestimated by our predictions 
may be due to the uncertainties affecting the 
spectro-photometric models used to determine the observational $M_{*}$. In this case, we plot the stellar 
masses obtained by Maiolino et al. (2008) by adopting the spectral synthesis models of 
Bruzual \& Charlot (2003). 
Maiolino et al. (2008) published also stellar masses obtained with the Maraston (2005) 
spectral synthesis models, in general producing stellar masses lower by $\sim 0.1$ dex or less.  
The use of these stellar masses would reduce the discrepancy between our predictions and the observations. \\
In the SFR-Mass plot, most of the observations, 
plotted with their error bars, 
overlap with predictions for galaxies with intermediate dust surface densities. 
An important caveat concerning the observed SFRs we should keep in mind is that, 
in most  cases, it is likely that the observed SFRs may represent underestimations to the 
real values. In general, in observational studies of high-redshift galaxies, 
the SFRs and the ages are very uncertain parameters. 
The SFRs are determined from the UV luminosity by 
assuming that the age of each galaxy is larger than 50 Myr. 
By removing this constraint, the estimated SFRs may be higher by a factor of 3 for the youngest objects. 
As a consequence, all the points would move upwards, towards regions characterized 
by lower dust surface density, hence less dust-obscured.\\
The evolution of the ellipticals is very quick within the first 0.2 Gyr (see fig.~\ref{SFR_ell}), 
such as the accumulation of a large dust mass. 
Fig.~\ref{Mz_dust} indicates that, if the AMAZE sample at z=3.5 is composed 
primarily by the progenitors of ellipticals, most of these galaxies may remain 
observable for times $\le 0.2$ Gyr, considering also the uncertainties 
in the observed SFRs. After 0.2 Gyr, the bulk of galaxies 
may become
heavily dust-obscured and cannot be detected by current surveys in the optical-UV bands. \\
A population of extremely dust obscured and vigorously star-forming galaxies 
is represented by the SCUBA galaxies (Clements et al. 2008), 
very luminous in the far infrared and sub-millimetric bands owing to 
massive dust reprocessing, but intrinsically faint in the optical band. 
Schurer et al. (2009) have used the same elliptical 
models described in this work to compute the photometric 
properties of galaxies,  by taking into account 
both the chemical and the spectro-photometric evolution of silicate and carbon dust 
grains. Schurer et al. (2009) showed that the general shape of the spectral energy distribution 
and the observed amount
of dust reprocessing of SCUBA galaxies are correctly reproduced by means of our models.\\
In summary, for ellipticals we predict metallicities larger than the observation owing to the fact that the evolution 
of our galaxies is very fast. A possible way to reduce the predicted metallicities could invoke longer infall times
for the ellipticals. A longer infall timescale would cause in each galaxy a longer star formation period and lower star formation rates, 
consequently a slower growth of the metallicity versus time. 
Another possibility could invoke stronger outflows in the elliptical models, or 
metal-enhanced ouflows, where the metallicity of the ejected matter is higher than the one of the ISM. However, 
this would require the introduction of additional parameters in the elliptical model.

\subsection{MZ relation for various elements and interstellar abundance ratios}
\label{alpha_fe}
\begin{figure*}
\centering
\epsfig{file=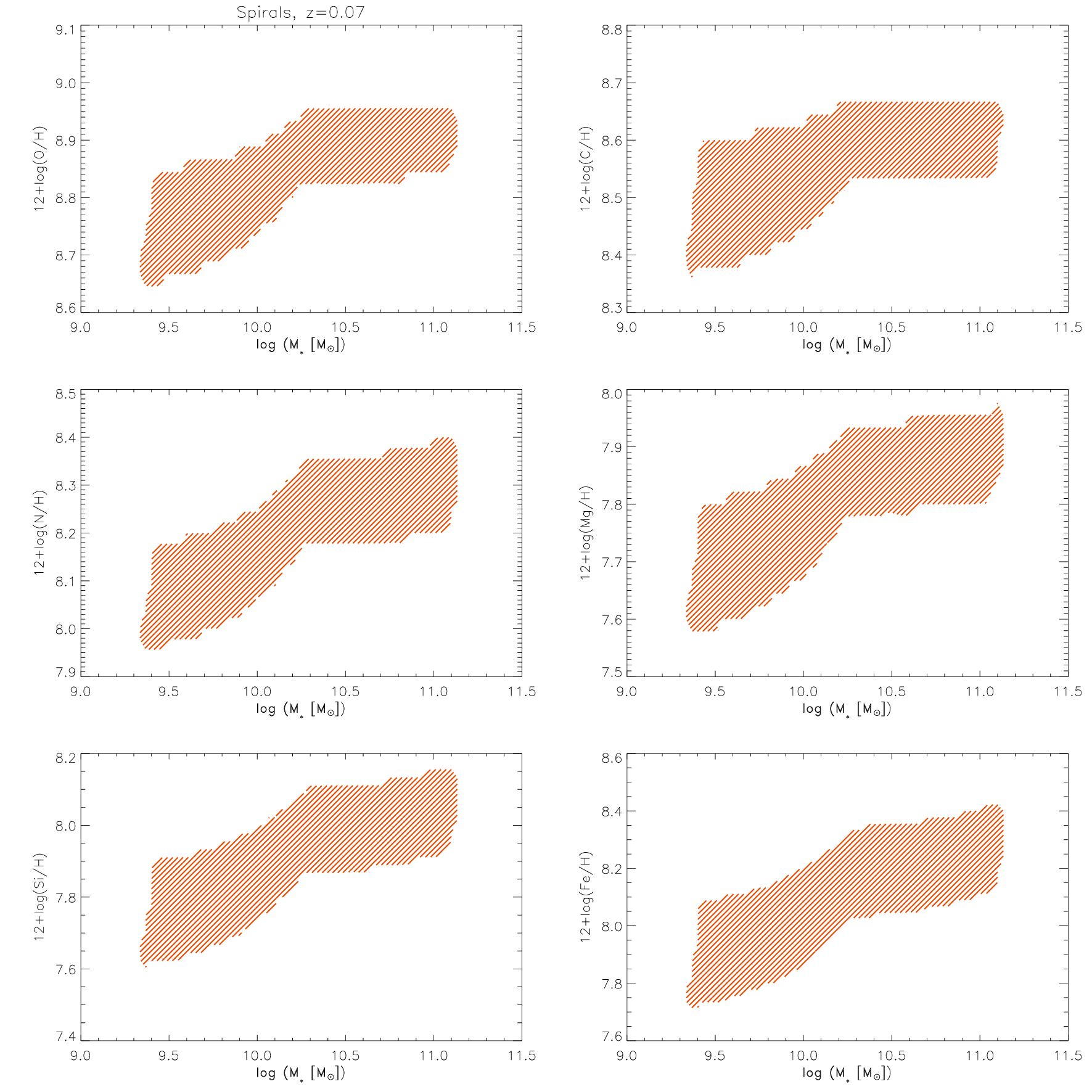, height=16cm,width=16cm}
\caption[]{Predicted MZ relations for various elements in spiral galaxies. 
Here we have assumed $z_f=3$ and $\Delta_t=10$ Gyr. 
\label{Mz_elements} }
\end{figure*}

In Fig.~\ref{Mz_elements}, we show our predictions for the MZ relations 
for spiral galaxies, considering on the y axis various chemical elements: 
O, C, N, Mg, Si, Fe. Here we have assumed $z_f=3$ and $\Delta_t=5$ Gyr. 
These elements are produced by different stellar sources and restored into the ISM 
on different timescales (see Matteucci 2001; Calura \& Matteucci 2006b). 
C and N are produced mainly by low and intermediate mass stars of 
mass $<8 M_{\odot} $. Mg, Si and O are produced by massive stars, dying as Type II SNe. 
However, Mg and Si are produced also by Type Ia SNe in a non negligible amount. 
Finally, the bulk of Fe is produced by Type Ia SNe. 
The maximum dispersion of the predicted MZ relations (i.e. the dispersion at stellar masses 
$10^{9.2} - 10^{9.3} M_{\odot} $ )
reflect the different production timescales of the elements and the lifetimes of the stars where they are synthesised. 
The smallest dispersion is of $\sim 0.2$ dex, visible for O.  
Dispersions of $\sim 0.22-0.25$ dex are predicted for C, N and Mg. 
The largest dispersions are predicted for the elements Si ($\sim$ 0.3 dex) and Fe ($\sim$ 0.4 dex), produced in considerable amounts 
by Type Ia SNe. This is related to the fact that the 
lifetimes of stars exploding as Type Ia SNe span a very large range, from $0.03$ Gyr for 
the highest masses ($M=8 M_{\odot}$ ) up to a Hubble time for the lowest mass stars. 
The reason for the larger scatter in the elements produced by Type Ia SNe  
with respect to the $\alpha$ elements can be understood by looking at Fig.~\ref{SFR_spi}. While the Type II SN rate, scaling with the rate of production of the 
$\alpha$ elements, presents a peak at early times and  then decreases progressively, the Type Ia SN rate presents a monotonic increase with cosmic time, 
hence the massive production of the Fe-peak elements lasts for a longer time than the $\alpha$-element production. 

\begin{figure*}
\centering
\epsfig{file=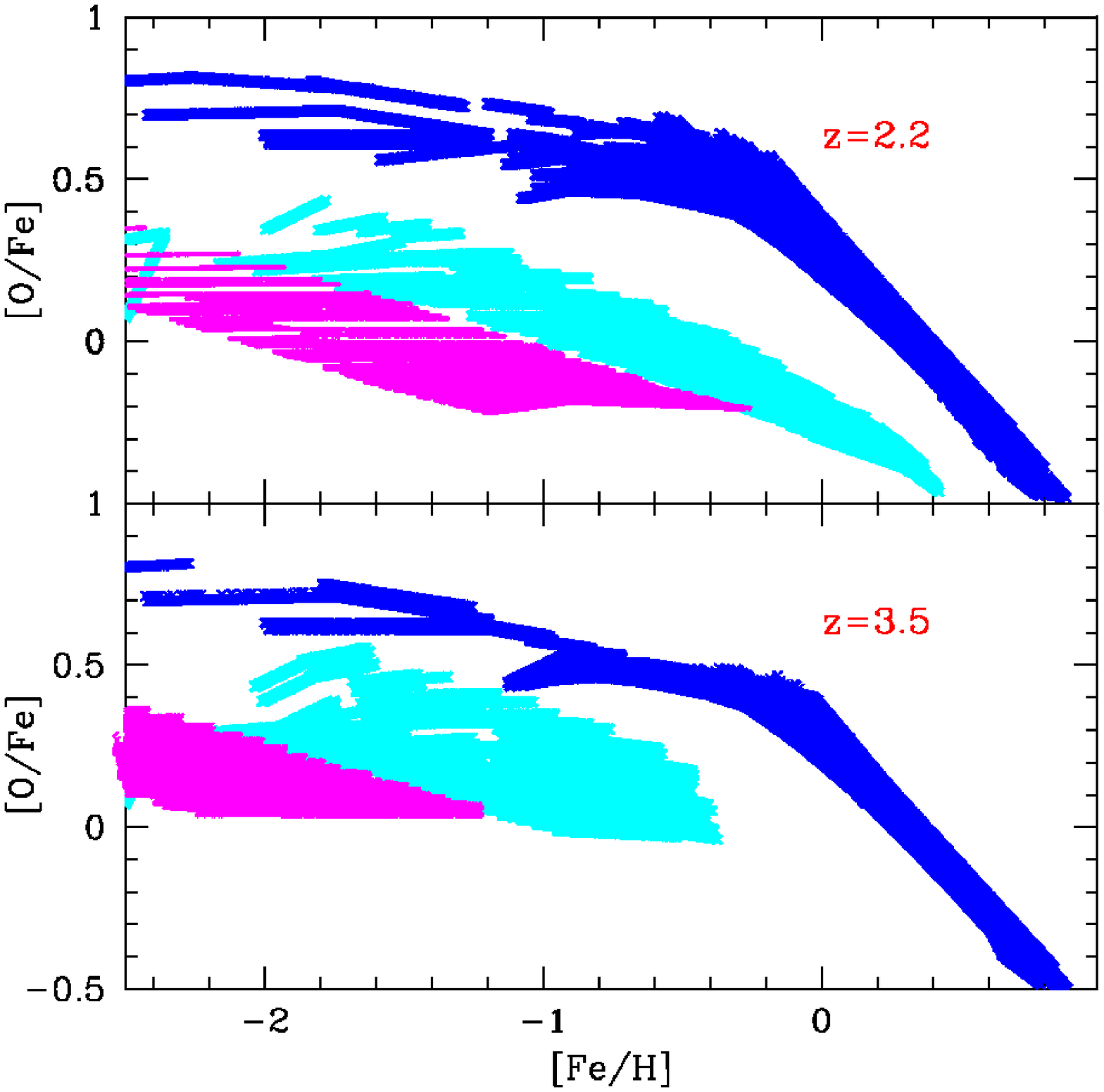, height=16cm,width=16cm}
\caption[]{Predicted [O/Fe] vs [Fe/H] at $z=3.5$ (lower panel) and  at $z=2.2$ (upper panel) 
for elliptical galaxies (blue areas), spirals (cyan areas) and irregulars (magenta areas). 
Here we have assumed $z_f=3$ and $\Delta_t=10$ Gyr for spirals and irregulars and $z_f=3$ and $\Delta_t=3$ for 
ellipticals. 
\label{O_Fe} }
\end{figure*}

\subsection{The importance of abundance ratios}
In Sect.~\ref{evol}, we have seen that, in the MZ, O/H vs SFR and SFR vs Mass plots, 
the predictions for ellipticals and spirals overlap, thus making difficult the interpretation of the observational data. 
A way to overcome this issue  
is to consider diagrams which involve abundance ratios between two different chemical 
elements. 
The abundance ratios between two chemical 
elements synthesized on 
different timescales 
can be used as "cosmic clocks", providing us with information on the relative 
roles of various stellar sources in the chemical enrichment of the interstellar medium (Matteucci 2001). 
In particular, the study of the [$\alpha$/Fe]\footnote{All the abundances between two different elements X and Y 
are expressed as $[X/Y]=log(X/Y)-log(X/Y)_{\odot}$, where  $(X/Y)$ and $(X/Y)_\odot$ are 
the ratios between the mass fractions of X and Y in the ISM and in the sun, respectively. 
We use the set of solar abundances as determined by Grevesse at al. (2007).} is of major importance, 
owing to the difference in the timescales for $\alpha$-elements and Fe production. \\
In Fig.~\ref{O_Fe}, we show the predicted [O/Fe] plots computed 
at $z=3.5$ and at $z=2.2$ for ellipticals, spirals and irregulars. 
As one can see, in both plots the three morphological 
types occupy very distinct  regions. This effect is primarily due to the 
star formation histories: ellipticals galaxies form quickly, by means of intense starbursts. 
For this reason, at low metallicity (e.g. [Fe/H]$\le -1$)
the enrichment of their ISM is dominated by $\alpha$-elements, produced by short-living massive stars. 
Owing to their short and strong 
starbursts, ellipticals reach high metallicities very quickly by means of Type II SNe, thus having high [$\alpha$/Fe] ratios for a large range of [Fe/H].
On the other hand,  spirals and irregulars have lower star formation 
rates and more prolonged star formation histories. For this reason, they reach 
high metallicities at later times, when the contribution of Type Ia SNe, producing 
the bulk of Fe and exploding on long timescales, becomes important. 
In principle, if measures of interstellar Fe abundances in high-redshift star-forming galaxies were accessible, 
thanks to our chemical evolution models, it would be possible to have important hints on 
the morphology of these objects. \\
The measure  of metal abundances in high-redshift star forming galaxies 
has become accessible in the recent years, thanks to the use of  
 10 m telescopes for spectroscopic studies, for which 
bright Lyman-break galaxies observed at $z>2$ are particularly valuable tools. 
Unfortunately, in these objects the measure of abundances other than O are hampered by weak 
emission lines, depletion uncertainties and unknown ionization corrections (Pettini 2004; Leitherer 2005). 
More promising objects for interstellar abundance ratio studies are Damped-Lyman Alpha galaxies, 
observed in absorption in the spectra of background Quasars. 
For these systems, the use of high-resolution spectrographs 
offers the access to weak interstellar lines, which lead to precise 
determination of interstellar abundances for various elements (Dessauges-Zavadsky et al. 2007).  
Recently, an attempt to study the possible existence of a mass-metallicity relation also in Damped 
Lyman Alpha systems has been suggested in a pioneeristic work by Ledoux et al. (2006). 
The interpretation of these data by means of models, such as the ones presented in this paper,  
will represent the next step in cosmic chemical evolution studies.

\section{Conclusions}
In this paper, we have performed a study of 
the evolution of the MZ relation in galaxies  
of different morphological types, and compared our results with the most recent 
observational data. 
In particular, we have used various observations achieved by several 
different authors, 
adopting a common metallicity calibration suited to 
our study, where the measured metallicities of the 
galaxies, observed at redshifts from 0.07 to 3.5, 
span a large range of values. 
We have used chemical evolution models for ellipticals, spirals and irregular galaxies. 
The models used here are able to reproduce the main properties of 
local galaxies. For each morphological type, 
we have used three baseline models of different present-day 
stellar masses.   
We have assumed that galaxy morphologies
do not change with cosmic time. 
We have developed a method to take into account a possible spread in the epochs of galaxy formation, 
namely in the times at which different galaxies start forming stars. 

Our main results can be summarized as follows. \\
1) Previous papers interpreted the observed MZ-relation in terms of the Simple Model of galactic chemical 
evolution. 
In this model, the main quantity linked  to the galactic metal content is the ``effective yield''  $y_Z = Z/ln(\mu^{-1})$, where 
$\mu$ is the gas fraction. 
In reality, galaxies will suffer infall/outflow and the true yield will be lower than the effective one.
The strongest assumption about the Simple Model is the instantaneous recycling approximation, 
and if elements different than O would be 
considered, such as N or Fe, this hypothesis would be very poor, since both N and Fe are produced  
mainly by low and intermediate mass stars on long timescales. For O, the IRA can still be acceptable.
Therefore, interpreting the MZ-relation in terms of the Simple Model has led several authors to conclude 
that larger galaxies should have a larger true yield than less massive ones. This can be achieved in several 
ways: i) by varying the IMF, ii) by decreasing the importance of outflows in more massive galaxies and iii) 
by decreasing the importance of infall in more massive galaxies. In this paper, we have shown that 
a MZ-relation naturally arises if less massive galaxies experience a lower star formation efficiency 
(i.e. the star formation rate per unit mass of gas), 
irrespective of the galactic morphological type. Galactic winds in less massive galaxies are therefore 
not necessary to explain the MZ-relation. \\
From the observational point of view, it 
is not clear if the effective yield decreases or increases with galactic mass (see Tremonti et al. 2004; Erb et al. 2006a). 
New data on the redshift evolution of the gas fractions and of the effective yields have recently become available 
(Calura et al. 2008b; Mannucci et al. 2009). 
A forthcoming paper will be focused on the study of the redshift evolution of both the gas fractions and of effective yields, considering 
galaxies of various morphological types. \\
2) At $z=0.07$, the slope and the zero point of the observed 
MZ relation are affected by uncertainties related to the use of different metallicity calibration methods. 
Our predictions, obtained with 
models able to reproduce a large set of independent chemical evolution constraints for spirals and irregulars, 
may  be useful 
to constrain the best calibration method. 
By means of our predictions, we suggest 
the best calibrations methods to be used by observers to derive gas metallicities. 
Our results are consistent with the MZ relations obtained by means of the calibration 
of Kewley \& Dopita (2002), Denicolo et al. (2002) and Pettini \& Pagel  (2004). \\

3) From  the redshift evolution of the MZ relation, 
we have seen how galaxies of various morphological types 
occupy different regions in the MZ plots at various redshifts. However, in these plots, 
in some cases, predictions for different galaxies tend to overlap. 
For this reason, the MZ plot alone cannot be used to constrain the morphology 
of the observed galaxies. \\

4) To constrain the morphological type of the observed galaxies, 
a very helpful method is to consider, beside the MZ plot, 
O/H vs SFR and SFR vs mass diagrams. 
We have tested how the adoption of different values for 
the parameters used in this study, $z_f$ and $\Delta_t$, 
influences the predictions at various redshifts. 
At $z=3.5$, $z=2.2$ and $z=0.7$, our models can reproduce both 
the metallicities and the SFRs of the observed galaxies by assuming 
for all galaxies 
an average redshift of formation of $z_f=3$. 
An age dispersion of $\Delta_t=3$ Gyr for ellipticals, compatible with the values 
coming from photometric studies of the Colour-magnitude relations of early-type galaxies 
in clusters and in the field, $\Delta_t=5$ Gyr for spirals and $\Delta_t=10$ irregulars, allows 
us to reproduce most of the observational constraints considered in this work.

5) The contribution of each type of star 
forming galaxy to the MZ relation is a fuction of redshift. 
At $z=3.5$, the observed stellar masses, metallicities and SFRs are reproduced by our models 
for elliptical galaxies. 
At $z=2.2$, the observed metallicities and SFRs indicate that the galaxies are likely to 
represent a morphological mix, composed partly by spirals (or proto-spirals) and partly by 
(proto-)ellipticals.
At $z=0.7$, we slightly underestimate the observed metallicities. 
Furthermore, we tend to overestimate the observed SFRs, which, plotted against the stellar mass $M_{*}$, 
show a weaker correlation than our predictions. 
At $z=0.07$, we reproduce the shape of the observed MZ relation. However, the dispersion 
of the data is larger than what our predictions indicate. The discrepancies between predictions and observations 
at $z=0.7$ and $z=0.07$ could be due to the fact that, in our chemical 
evolution models, the star formation histories do not take into account stochastic events such as
episodic starbursts. Such events could increase the dispersion of the observed metallicities and make the 
real star formation histories more irregular than the predicted ones.
In any case, this does not represent a major concern since, as shown in Kewley \& Ellison (2008), 
also the dispersion depends on the calibration. 
Our predicted SFRs are compatible with the observed values of the SDSS galaxies, although this sample 
includes also objects with very high SFRs, up to $log(SFR/M_{\odot}/yr)=1.8-2$, which we can not reproduce and which are sometimes 
even higher  than the values observed at $z=0.7$ in the GEMINI  Deep Deep Survey (Juneau et al. 2005). A possible reason for such high SFRs could 
be due to overestimated dust attenuation in the SDSS sample (J. Brinchmann, private communication). \\

6) At $z=3.5$, the majority of the predicted metallicity 
values for the ellipticals are  higher than the observed ones by $\sim 0.5$ dex. 
This difference is larger than the offset due to the use of different calibrations. 
By computing the dust surface mass densities for our elliptical models, 
we show that, 
if the AMAZE sample is composed mainly by 
the progenitors of local ellipticals, 
most of these galaxies may remain 
observable for times $\le 0.2$ Gyr, considering also the uncertainties 
in the observed SFRs. After 0.2 Gyr, the bulk of galaxies 
may become 
heavily dust-obscured and hardly detectable by current surveys in the optical-UV bands. \\

7) Our study of the MZ relation for various elements in spirals 
shows that the predicted dispersion at the lowest stellar masses 
depends on the lifetimes of the stellar progenitors 
producing the elements. The largest dispersions in the MZ plot are predicted 
for Si and Fe, which are produced by Type Ia SNe in considerable amounts. \\

8) An analysis of  the predicted [O/Fe] plots computed 
at $z=3.5$ and at $z=2.2$ for ellipticals, spirals and irregulars 
shows that, if interstellar Fe abundances in high-redshift star-forming galaxies were measurable, 
thanks to our chemical evolution models, it would be possible to have important hints on 
the morphology of these objects. \\
Unfortunately, interstellar Fe abundance measures from emission lines are not 
achievable by means of present-day instruments. 
However, the determination  of 
the iron abundance of star-forming galaxies at high redshift, as performed  by Halliday et al. (2008), 
which use the Rix et al. (2008) Fe absorption line, 
could represent an important result in this regard. The combination of these abundances 
with measures of the interstellar O abundance would provide the observables required by our study.\\
It is also worth to note that Fe abundances can be obtained in absorption in gas-rich galaxies observed in the spectra 
of high-redshift QSOs, such as Damped-Lyman Alpha galaxies, together with 
the abundances of various other elements. A seminal work of Ledoux et al. (2006) points 
towards the existence of a possible mass-metallicity relation also in DLAs. In the future, 
this issue, which  needs to be confirmed by further observations,  
will be an interesting subject for forthcoming theoretical chemical evolution studies.\\

\begin{acknowledgements}
G. De Lucia is acknowledged for interesting discussions. \\
This work was partially supported by the Italian Space Agency
through contract ASI-INAF I/016/07/0.
F.C. and F. M. acknowledge financial support from PRIN2007, Prot.2007JJC53X\_001.
C.C. acknowledgs financial support from Swiss National Science Foundation (SNF).
\end{acknowledgements}

\end{document}